\documentstyle[fund2]{article}

\renewcommand{\textwidth}{16.5cm}

\newcommand{\Ne}{N_e}

\newcommand{\Qbare}{Q_b}
\newcommand{\Qstar}{Q^{*}}

\newcommand{\kfirst}{k^{(1)}}
\newcommand{\ksecond}{k^{(2)}}
\newcommand{\kbulk}{k^{{\small bulk}}}
\newcommand{\kbulkradical}{k^{\small bulk}_{\small rad}}

\newcommand{\kone}{k^{(1)}}
\newcommand{\ktwo}{k^{(2)}}

\newcommand{\Pmany}{P_m}
\newcommand{\Ptwo}{P_2}

\newcommand{\Rt}{{\cal R}_t}
\newcommand{\Rdot}{\dot{\Rt}}
\newcommand{\Rtdot}{\dot{\Rt}}

\newcommand{\Rtbulk}{{\cal R}_t^{\small bulk}}
\newcommand{\Rtdotbulk}{\dot{\Rt}^{\small bulk}}

\newcommand{\Gt}{G_t}
\newcommand{\Gtone}{G_t^{(1)}}
\newcommand{\Gttprime}{G_{t-\tprime}}
\newcommand{\Sd}{S^{(d+1)}}
\newcommand{\Sone}{S^{(1)}}
\newcommand{\Ima}{I_m^A (t)}
\newcommand{\Imb}{I_m^B (t)}

\newcommand{\ta}{t_a}
\newcommand{\th}{t_h}
\newcommand{\tstartwo}{t_{2}^{*}}
\newcommand{\tstarmany}{t_{m}^{*}}

\newcommand{\tl}{t_l}
\newcommand{\tprime}{{t}'}
\newcommand{\Tl}{T_l}
\newcommand{\te}{t_e}
\newcommand{\tb}{t_b}
\newcommand{\xt}{x_t}
\newcommand{\xa}{x_A}
\newcommand{\xb}{x_B}

\newcommand{\ra}{{\bf r}_A}
\newcommand{\rat}{{\bf r}_A^T}

\newcommand{\raprime}{\ra'}
\newcommand{\ratprime}{{\rat}'}
\newcommand{\rb}{{\bf r}_B}
\newcommand{\rbt}{{\bf r}_B^T}
\newcommand{\rbprime}{\rb'}
\newcommand{\rbtprime}{{\rbt}'}
\newcommand{\r}{{\bf r}}
\newcommand{\rt}{{\bf r}^T}
\newcommand{\rprime}{\r'}
\newcommand{\rtprime}{{\rt}'}

\newcommand{\bv}{{\bf v}}
\newcommand{\bu}{{\bf u}}
\newcommand{\ns}{n^s}

\newcommand{\nainf}{n_{A}^{\infty}}
\newcommand{\nbinf}{n_{B}^{\infty}}

\newcommand{\nofinfinity}{n(\infty)}

\newcommand{\na}{n_A}
\newcommand{\nb}{n_B}
\newcommand{\nas}{n_A^s}
\newcommand{\nbs}{n_B^s}

\newcommand{\rhoabs}{\rho _{AB}^{s}}
\newcommand{\rhoab}{\rho_{AB}}
\newcommand{\rhoaa}{\rho_{AA}}
\newcommand{\rhoaas}{\rhoaa^s}
\newcommand{\rhobbs}{\rhobb^s}
\newcommand{\rhobb}{\rho_{BB}}
\newcommand{\rhoabb}{\rho_{ABB}}
\newcommand{\rhoaba}{\rho_{ABA}}
\newcommand{\rhoaab}{\rho_{AAB}}

\newcommand{\rhobab}{\rho_{BAB}}
\newcommand{\rhobba}{\rho_{BBA}}

\newcommand{\psinudag}{\psi_{\nu}^{\dag}}
\newcommand{\psimudag}{\psi_{\mu}^{\dag}}
\newcommand{\psinu}{\psi_{\nu}}
\newcommand{\psimu}{\psi_{\mu}}
\newcommand{\psia}{\psi_A}
\newcommand{\psib}{\psi_B}
\newcommand{\psibdag}{\psi_B^{\dag}}
\newcommand{\psiadag}{\psi_A^{\dag}}

\newcommand{\phiB}{\phi_B}

\newcounter{fignumber}
\begin{document}

%\bibliographystyle{benaip}

%****************************** title page *****************************************

\renewcommand{\thepage}{}

\titleben{\LARGE \bf Interfacial Reaction Kinetics}

\author{\Large 
BEN O'SHAUGHNESSY\ $^{1}$ \ and \ DIMITRIOS VAVYLONIS\ $^2$ \\ 
}

\maketitle

\ \\ \newline
{\large \center
$^1$ Department of Chemical Engineering \\
Columbia University\\ 
500 West 120th Street \\
New York, NY 10027 \\
e-mail: bo8@columbia.edu\\
} 

\vi

{\large \center
$^2$ Department of Physics \\ 
Columbia University \\
538 West 120th Street \\
New York, NY 10027 \\
e-mail: dvav@phys.columbia.edu\\
}

\vii

\  \newline

%Submitted to: {\em  European Physical Journal B}\\

\ignore{
Corresponding author:
\parbox[t]{4in}{
Ben O'Shaughnessy

Fax: ++(212)-854-3054

email: beno@chem352.cheme.columbia.edu
}
}% end \ignore

\pagebreak

%****************************** abstract page *****************************************

\pagenumbering{arabic}

\large

\section*{ABSTRACT}

We study irreversible A-B reaction kinetics at a fixed interface separating two
immiscible bulk phases, A and B.  Coupled equations are derived for
the hierarchy of many-body correlation functions.  Postulating
physically motivated bounds, closed equations result without the need
for ad hoc decoupling approximations.  We consider general dynamical
exponent $z$, where $x_t\sim t^{1/z}$ is the rms diffusion distance after
time $t$.  At short times the number of reactions per unit area,
$R_t$, is {\em 2nd order} in the far-field reactant densities
$n_A^{\infty},n_B^{\infty}$.  For spatial dimensions $d$ above a critical value
$d_c=z-1$, simple mean field (MF) kinetics pertain, $R_t\sim Q_b t
n_A^{\infty} n_B^{\infty}$ where $Q_b$ is the local reactivity.  For low
dimensions $d<d_c$, this MF regime is followed by 2nd order diffusion
controlled (DC) kinetics, $R_t \approx x_t^{d+1} n_A^{\infty} n_B^{\infty}$,
provided $Q_b > Q_b^* \sim (n_B^{\infty})^{[z-(d+1)]/d}$.
Logarithmic corrections arise in marginal cases.  At long times, a
cross-over to {\em 1st order} DC kinetics occurs: $R_t \approx x_t
n_A^{\infty}$.  A density depletion hole grows on the more dilute A side.
In the symmetric case ($n_A^{\infty}=n_B^{\infty}$), when $d<d_c$ the long time
decay of the interfacial reactant density, $n_A^s$, is determined by
fluctuations in the initial reactant distribution, giving $n_A^s \sim
t^{-d/(2z)}$.  Correspondingly, A-rich and B-rich regions develop at
the interface analogously to the segregation effects established by
other authors for the bulk reaction $A+B\rightarrow\emptyset$.  For $d>d_c$
fluctuations are unimportant: local mean field theory applies at the
interface (joint density distribution approximating the product of A
and B densities) and $n_A^s \sim t^{(1-z)/(2z)}$.  We apply our results
to simple molecules (Fickian diffusion, $z=2$) and to several models
of short-time polymer diffusion ($z>2$).

\vii

PACS numbers: 05.40.+j, 68.45.Da,  82.35.+t\\

\ignore{
  \begin{benlistdefault}
    \item [05.40.+j]
 (Fluctuation phenomena, random processes, and Brownian Motion)
    \item [68.45.Da]
 (Diffusion; interface formation) 
    \item [82.35.+t]
 (Polymer reactions and Polymerization)
  \end{benlistdefault}
} % end \ignore

\pagebreak

%****************************** PAPER  *****************************************

\section{Introduction}

In a large class of chemically reacting systems, irreversible
bimolecular reactions occur at a permanent interface separating two
bulk phases.  Reactive molecules in one phase are able to react with
molecules in the other phase only; hence reaction events can occur
within the limits of the narrow interfacial region only.  A number of
technologically important examples
\citeben{astarita:book,doraiswamysharma:book} entail small molecules
reacting at liquid-liquid, liquid-gas or liquid-solid interfaces.  In
the present study we address interfaces which are fixed in space and
do not broaden as reactions proceed; the two bulk phases do not mix
with one another.  However, the physics we will explore may provide
insight to the very different problem of non-stationary reactive
fronts where chemical reactions occur at a moving and possibly
broadening interface separating miscible phases
\citeben{galfiracz:aplusbfront,%
cornelldrozchopard:aplusbfront_fluc,%
cornelldroz:aplusb_front_prl,%
araujo:anomalies_aplusb_front,%
taitelbaum:aplusbfront,%
leecardy:aplusb_front,%
howardcardy:aplusb_front_rg,%
barkema:aplusb_front_1d,%
kookopelman:aplusbfront_expmt}.

Another important class of reactive interfacial systems involves
macromolecules.  Of particular technological significance is the
process of reactive blending
\citeben{okamotoinoue:react_proc,sundararajmacosko:utm} where the compatibilization of
two immiscible polymer melts A and B is assisted by attaching reactive
groups to a certain fraction of the chains. The A-B copolymers
generated by reactions, which can occur at the A-B melt interface
only, serve both to reinforce the interface
\citeben{kramer:strength_iface_faraday,gersappe:bind_blends_copolym}
and to promote the mechanical mixing of the two
melts \citeben{sundararajmacosko:utm,milnerxi:copol_mixing}.

The manner in which reaction kinetics are modified by the presence of
an interface is a fundamental issue within the general field of
chemical reaction kinetics.  Despite this, and despite the numerous
applications such as those mentioned above, no complete and systematic
theory exists.  We emphasize that each reaction event necessitates the
simultaneous arrival, at the same location within the interface, of two
molecular species A and B, one from each bulk phase.  (This is very
different to the problem \citeben{carslawjaeger:book,ben:adsorption}
of a single bulk adjacent to a homogeneous ``reactive interface''
where each ``reaction'' event, e.g. the irreversible adsorption of a
molecule onto a solid surface, requires the arrival of only {\em one}
molecule at the interface.)  The interfacial reaction kinetics which
are the subject of the present paper were theoretically studied for
the case of small molecules by Durning and O'Shaughnessy
\citeben{ben:reactiface}, and the end-functionalized polymer case by
O'Shaughnessy and Sawhney
\citeben{ben:reactiface_pol_letter,ben:reactiface_pol} and Fredrickson
\citeben{fredrickson:reactiface_prl}.  These theories in fact described
a certain short time regime only, for systems where the reacting
species are dilute in an unreactive background.  Fredrickson and Milner
\citeben{fredricksonmilner:reactiface_timedept} argued that at later
times different kinetic behaviors onset, with forms dependent on
reactive species concentration.

In this paper we develop a near-exact theory of irreversible
interfacial reaction kinetics.  We calculate time-dependent reaction
rates as a function of density and local reactivity of the reactive
species.  In addition, density profiles on either side of the
interface are determined.  A short version of the present manuscript
has appeared
\citeben{ben:fund_letter_toappear}.  Our framework is quite general in
terms of the diffusive dynamics of the reactive species, as defined by
the dynamical exponent, $z$\ :
%_______________________________________________________________________
                                                \begin{eq}{xt}
\xt = a\, \power{t}{\ta}{1/z} \comma
                                                                \end{eq}
%-----------------------------------------------------------------------
where $\xt$ is the rms displacement of a reactive group after time
$t$.  Here $a$ is the linear dimension of the reactive species A and
B, and $\ta$ is the diffusion time corresponding to $a$.  Thus,
setting $z=2$ in our results yields reaction rates for small molecules
obeying Fick's diffusion law.  As a second example, if one seeks the
short time reaction kinetics of small reactive groups attached to
polymer chains in the melt, then appropriate values would be $z=4$ or
$z=8$, depending on the time regime and degree of entanglement
\citeben{doiedwards:book,gennes:book}.

This study will always assume the interface is clean.  Thus we ignore
effects associated with accumulation of A-B reaction products at the
interface whose presence may eventually diminish reaction rates
\citeben{ben:reactiface_pol_letter,ben:reactiface_pol,fredricksonmilner:reactiface_timedept}.

\subsubsection*{Bulk Kinetics (A+B $\gt$ 0): a Brief Review}

Before attacking the present interfacial problem, it is helpful to
consider first the analogous and somewhat simpler problem in the {\em
bulk}, where irreversible bimolecular reactions between A and B,
generating inert products, can occur anywhere within a single bulk
phase.  There are no A-A or B-B reactions.  For the small molecules
case, $z=2$, many well-established results exist (``$A+B\gt 0$'')
\citeben{ovchinnikovzeldovich:segregation,%
toussaintwilczek:segregation,%
meakinstanley:aplusb_fractal,%
kangredner:segregation,%
leecardy:ab_rg}.
Suppose A and B have equal diffusivities and are initially uniformly
distributed with equal densities $n(0)$ within some solvent.  (The
case $n(0)=1/a^d$ would correspond to every molecule being reactive;
generally $n(0)\le 1/a^d$.)  Reactions are now ``switched on'' at
$t=0$.  Then, whenever an A and a B particle collide (i.e. approach to
within distance of order $a$ of one another) they react irreversibly
with probability $\Qbare$ per unit time, where $\Qbare$ is the local
reactivity.  For simplicity, let us confine our bulk discussion to
``infinitely'' reactive particles; reaction then occurs every time an
A-B pair collides.  This corresponds to setting
\citeben{ben:intersemi_all} $\Qbare=1/\ta$ (the effective local
reactivity cannot exceed the rate, $1/\ta$, at which diffusion can
bring two reactive species together).  Our discussion will consider a
general spatial dimensionality $d$.

What are the reaction kinetics in this bulk system?  What are the time
dependencies of the number of reactions per unit volume which have
occurred after time $t$, namely $\Rtbulk$, and the reactant density
$n(t)$?  The simplest guess is that mean field (MF) theory applies:
this amounts to assuming reactants are always distributed as in {\em
equilibrium}.  Hence the reaction rate equals the {\em equilibrium}
density of A-B pairs in contact, multiplied by $\Qbare$. Thus,
$\Rtdotbulk\equiv (d/dt)\Rtbulk= \Qbare a^d \ n^2(t) = (a^d/\ta)\
n^2(t)$.  Now this MF prediction is in fact valid only for
sufficiently large $d$ such that diffusion is effective in dissipating
reaction-induced non-equilibrium spatial correlations.  The maximum
number of A-B pairs which diffusion can have brought together by the
time $t$ increases as $\xt^d \twid t^{d/z}$; provided $d>z$, this is
sufficient to restore the depletion, which would arise in the two-body
correlation function, due to reactions as implied by the MF
prediction, $\Rtbulk\twid t$.  But for lower dimensions, $d<z$, since
diffusion cannot supply material fast enough to keep pace with this
reaction rate, equilibrium spatial correlations are destroyed: a
depletion hole of size $\xt$ grows in the A-B 2-body correlation
function.  Reaction kinetics are then very different; for short times
$\Rtbulk\approx \xt^d n^2(0)$ is the number of reactive pairs
initially within diffusive range of one another, \ie whose initial
separations were less than $\xt$.  In summary, for times short enough
that the relative density drop is small, $n(t)\approx n(0)$, we have
%_______________________________________________________________________
                                                \begin{eq}{bulk}
\Rtdotbulk = - {d n(t) \over dt} = \kbulk n^{2}(t) \comma\ \ \
\kbulk(t) \approx \casesbracketsii
{a^d/\ta}{d>z}
{d\xt^d/dt\twid t^{d/z-1}}{d<z}  \gap
\mbox{(short times).}
                                                                \end{eq}
%-----------------------------------------------------------------------
These are second order rate kinetics, with a rate constant, $\kbulk$,
which is time-dependent for low dimensions $d<z$.  

The two classes of kinetics in eq. \eqref{bulk} reflect the fact that
reactive groups explore space ``compactly'' in low dimensions
\citeben{doi:inter2,gennes:polreactionsiandii,ben:bulk_letter}: for $d<z$ it is
simple to show that (in the absence of reactions) the number of
collisions between an A-B pair, with some given initial separation,
increases for large times as $\twid t^{1-d/z}$.  Thus reaction is
inevitable by the time $t$ for any reactive pair initially separated
by $\xt$ or less.  By contrast, for $d>z$ space is explored in a
``non-compact'' or dilute fashion; with finite probability, the same
two particles may avoid collision as $t\gt\infty$.  This survival
probability is an increasing function of the initial separation.
Reaction is no longer inevitable between all pairs within diffusive
range, and MF theory applies \citeben{ben:interdil,ben:intersemi_all}.
For the interface problem, we will find a similar division between
compact and non-compact reaction kinetics, but now at a dimension
$d+1=z$.  Indeed, the short time interface kinetics turn out to be
analogous to those of a $d+1$-dimensional bulk problem
\citeben{ben:reactiface,ben:reactiface_pol_letter,ben:reactiface_pol}.

Eq. \eqref{bulk} describes the short time kinetics.  What happens at
very long times?  For $d>z$ one might anticipate the MF kinetics of
eq. \eqref{bulk} would continue indefinitely, implying $n\twid 1/t$
asymptotically.  In fact, for two-species A-B systems this is true
only for very high dimensions, $d>2z$.  This was demonstrated for
$z=2$ by Ovchinnikov and Zeldovich
\citeben{ovchinnikovzeldovich:segregation} and by Toussaint and
Wilczek \citeben{toussaintwilczek:segregation}.  These authors showed
that in lower dimensions fluctuations in the initial density
distribution determine the asymptotic form of $n(t)$.  Their argument,
generalized to arbitrary $z$, is roughly as follows.  Consider a
portion of the reacting system of volume $\Omega$ and let $N_A(t),
N_B(t)$ be the number of unreacted A and B particles in this region at
time $t$.  Assuming random initial spatial distributions of A and B,
the initial fluctuations of $N_A(0)$ and $N_B(0)$, about their mean
value $n(0) \Omega$, will be of order $\sqrt{n(0) \Omega}$.  (Note
that $n(t)$ is the {\em mean} density after time $t$.)  Of the same
order will be the fluctuations in $\delta N_0\equiv N_A(0) - N_B(0)$,
the average value of which is zero.  As reactions proceed,
fluctuations will diminish.  However, since reaction events conserve
the difference between the number of A and B particles, fluctuations
in $\delta N_t\equiv N_A(t) - N_B(t)$ can decay by diffusion only.
Thus, if we consider small regions, $\Omega<\xt^d$, then by time $t$
the initial difference of order $\delta N_0$ has had sufficient time
to decay away due to diffusion.  But for large regions,
$\Omega>\xt^d$, the difference must be close to its original value,
$\delta N_t \approx \delta N_0$.  Roughly, then, in a region of volume
$\xt^d$, the total number of reactants cannot be smaller than a number
of the order of $\sqrt{n(0) \xt^d}$. It follows that the density
cannot decay faster than $n(t)\approx \sqrt{n(0)
\xt^d}/ \xt^d \twid \sqrt{n(0)} t^{-d/(2z)}$.  For $d<2z$, this is a
slower decay than the MF $t^{-1}$ prediction, and one concludes that
this diffusive relaxation of initial fluctuations then governs the
long time decay.  To summarize,
%_______________________________________________________________________
                                                \begin{eq}{stuck}
n(t\gt \infty) \twid \casesbracketsii
{t^{-1}}{d>2z}
{t^{-d/2z}}{d<2z}  
                        \gap (\mbox{bulk}).
                                                                \end{eq}
%-----------------------------------------------------------------------
For interfacial reactions, we will establish a rather similar long
time fluctuation-dominated decay of densities near the interface for
sufficiently small $d$.  Analogously to the bulk case, this is
accompanied by segregation of reactants into A-rich and B-rich domains
of size $\xt$ in the region adjacent to the interface.

\subsubsection*{Interfacial Kinetics: Scaling Arguments}

Let us turn now to the interface problem, shown schematically in fig.
\ref{iface}. We consider two $d$-dimensional bulk phases
containing, respectively, reactive species A and B with initial
densities $\nainf$ and $\nbinf$.  The reactants are of size $a\le h$,
where $h$ is the width of the thin ($d$-1)-dimensional interfacial
region which is the locus of all reaction events.  We assume A and B
have identical diffusion dynamics.  To begin, consider the symmetric
case $\nainf=\nbinf\equiv n$ and the infinitely reactive limit,
$\Qbare\gt 1/\ta$ (every A-B collision produces a reaction).

The short time reaction kinetics of this interfacial system are
analogous to those of a ($d+1$)-dimensional bulk problem.  To see this,
consider how many degrees of freedom are needed to specify the
``reaction rate'' for a single A-B pair.  One coordinate must specify
how far from the interface particle A lies, and similarly for B.  A
further $d-1$ coordinates must specify their relative location,
giving $d+1$ degrees of freedom
\citeben{ben:reactiface} in total.  That is, there are $d+1$ diffusive
degrees of freedom which must vanish in order that an A-B pair may
react.  These are the reaction conditions for a $d+1$-dimensional bulk
diffusion-reaction problem, and similar reasoning to that for the bulk
dictates that non-compact MF kinetics pertain for $d+1>z$, whilst for
$d+1<z$ the kinetics are of compact diffusion-controlled (DC) form.
Thus the reaction rate per unit interface area, $\Rtdot$, obeys {\em
2nd order} rate kinetics with a {\em 2nd order} rate constant $\ktwo$:
%_______________________________________________________________________
                                                \begin{eq}{brown-scroll}
\Rtdot = \ktwo n^{2} \comma\ \ \
\ktwo(t) \approx \casesbracketsii
{h a^d /\ta}{d+1>z}
{d\xt^{d+1}/dt\twid t^{(d+1)/z-1}}{d+1<z}  \gap
(\mbox{short times}\comma \Qbare\ta=1)
                                                                \end{eq}
%-----------------------------------------------------------------------
in complete analogy to eq.\eqref{bulk} for the bulk problem, but with
$d$ replaced by $d+1$.  The mean field result for $d+1>z$ follows
because in equilibrium there are $h a^d n^2$ A-B pairs in contact per
unit area of interface 
$^($\footnote{
A slight complication here, which will be addressed in section 6, is
that this result is modified when $z<d+1<z+1$; in that case $\ktwo
\approx h^{d+1}/\th$ where $\th\equiv \ta (h/a)^z$.}$^)$.  The DC compact
kinetics are determined by the small fraction (at short times) of A-B
pairs which were initially separated by less than $\xt$; for $d+1<z$
any such pair will definitely have reacted by time $t$.  The number of
such A-B pairs per unit interfacial area is $\xt^{d+1} n^2$ (see fig.
\ref{iface}).  Note that the dimensions of $\ktwo$ are
$x^{d+1}/t$, as appropriate to a ($d+1$)-dimensional bulk problem.
For the remainder of this paper we will refer to $d+1>z$ and $d+1<z$
as the ``non-compact'' and ``compact'' cases, respectively.

Consider the long time behavior now.  This is completely different to
the bulk.  In fact at long times the effective dimensionality of the
problem changes from $d+1$ to 1 and, moreover, the reaction kinetics
become of {\em first} order.  Let us first investigate the compact
case, $d+1<z$.  Consider an A particle that was initially within a
distance $l$ of the interface, where $l \equiv n^{-1/d}$ is the
typical separation between reactants.  By time $\tl \equiv \ta (n
a^d)^{-z/d}$, its exploration volume will typically contain one B
particle, in the other bulk, which was initially within $l$ of the A
particle.  Since $d+1<z$, reaction is certain.  It follows that for
times $t>\tl$ the interface becomes, in effect, ``perfectly
absorbing:'' almost every reactive species reaching it will suffer a
reaction.  Thus a density depletion hole develops at the interface
(see fig. \ref{profile}) and the reaction rate is limited by
diffusion to the interface, $\Rt\approx
\xt n$.  One concludes that long time reaction kinetics are now {\em
first} order, with a {\em first} order rate constant $\kone$ given by
%_______________________________________________________________________
                                            \begin{eq}{glass-in-the-wine}
\Rtdot = \kone n \comma \gap
\kone \approx {d\xt \over dt} \twid t^{1/z-1}\gap (t\gt \infty)\period
                                                                \end{eq}
%-----------------------------------------------------------------------
Notice that the dimensions of $\kfirst$ are $x/t$, as would be
appropriate to a one-dimensional bulk problem.  In correspondence to
the kinetics being first order, this DC regime is accompanied by a
growing hole of size $\xt$ in the {\em one-body} density ``correlation
function,'' \ie in the density field itself, $n(\r)$.  This is very
different to the hole, also of size $\xt$, which grew in the {\em
two-body} density correlation function for the {\em second} order
$d+1<z$ compact DC kinetics at short times, eq. \eqref{brown-scroll}.
For that regime, the density field itself was unchanged from
equilibrium.

What are the long time kinetics for the non-compact case, $d+1>z$?
The answer is: the same as for the compact case.  The only difference
is that the cross over from $d+1$ to 1-dimensional behavior no longer
occurs at $\tl$, but at a timescale we name $\tstarmany$.  In this
case we can estimate $\tstarmany$ using a mean field picture since the
early kinetics are MF.  Consider an A particle initially within $\xt$
of the interface, as in fig. \ref{tstar_many}.  After time $t$, it has
made $t/\ta$ ``steps,'' a fraction $(h/\xt)$ of which were within the
interfacial region where B particles are present at density $n$.  Thus
for each of these interfacial steps the probability that the A
particle was in contact with any B particle is $na^d$.  Hence the
total reaction probability, $\Pmany$, is 
%_______________________________________________________________________
                                                \begin{eq}{pmany}
\Pmany(t) \approx \paren{{h \over \xt} {t \over \ta}} \paren{n a^d}
 \approx {h \over a} \ n a^d \, (t/\ta)^{1-1/z}  
                                \gap(\Qbare=1/\ta)\period
                                                                \end{eq}
%-----------------------------------------------------------------------
Setting $\Pmany(\tstarmany)=1$, one obtains $\tstarmany/\ta = [a/(h n
a^d)]^{z/(z-1)}$.  Thus, for $t>\tstarmany$ any A particle within
diffusional range of the interface will definitely have reacted with
a B.  This is a many-body effect; by $\tstarmany$ any A near the
interface is bound to have reacted due to the mean reaction field
created by all of the B molecules.  We conclude that for large times a
density depletion hole develops also for the non-compact case,
following the same kinetics as eq. \eqref{glass-in-the-wine}.

So far we have considered ``infinitely'' reactive species,
$\Qbare\approx 1/\ta$.  Such local reactivities $\Qbare$ are realized
for radicals \citeben{beckwith:radical_rateconstants_book}, and in
certain other processes such as phosphorescence quenching
\citeben{mitahorie:review}.  However, these are very exotic exceptions
to the general rule: in virtually all practical situations $\Qbare$ is
tiny, $\Qbare \ta \lsim 10^{-6}$.  Indeed, for the vast majority of
reacting species, $\Qbare$ values are many orders of magnitude smaller
than $10^{-6}$
\citeben{denisov:rateconstants_book,bernasconi:rateconstants_book}.
It is essential, therefore, to establish how the picture we have
developed above is modified for finitely reactive systems.  For
non-compact cases, $d+1>z$, there is no qualitative change from the
kinetics of eq. \eqref{brown-scroll}: again, a short time 2nd order MF
regime, now with $\ktwo=\Qbare h a^d$, is followed at $\tstarmany$ by
a DC first order regime, but now the formula for $\tstarmany$ is
modified.  Notice that the expression for $\Pmany(t)$ in eq.
\eqref{pmany} is the mean number of collisions experienced by the A
particle; multiplying this by the reaction probability per collision,
$\Qbare \ta$, yields the total reaction probability for general
$\Qbare$. Defining $\Pmany(\tstarmany)\equiv 1$, we have
%_______________________________________________________________________
                                                \begin{eq}{tstarmany}
\Pmany(t) \approx (Q \ta) \ n a^d \, \power{t}{\ta}{1-1/z}  
          \comma \gap {\tstarmany\over\ta}= \power{1}{Q\ta \, n a^d}{z/(z-1)}
          \comma\gap     Q \equiv \Qbare \, {h\over a}
                                                                \end{eq}
%-----------------------------------------------------------------------
where $Q$ emerges as an {\em effective} local reaction rate
coarse-grained over the interface width $h$.

In the compact case, $d+1<z$, kinetics are more fundamentally modified
by finite reactivity.  For $\Qbare\ta =1$ we have seen an initial 2nd
order DC regime followed at $\tl$ by 1st order DC kinetics.  But for
$\Qbare \ta<1$, a new MF regime appears at early times.  Consider an
A-B pair near the interface whose members are initially closer than
$\xt$ to one another, as in fig. \ref{iface}. By time $t$, A has
taken $(t/\ta)(h/\xt)$ steps in the interface.  For a fraction
$(a/\xt)^d$ of these, B was in contact with A since B is equally
likely to be anywhere within its exploration volume $\xt^d$.  Thus,
the 2-body reaction probability for this pair is given by
%_______________________________________________________________________
                                                \begin{eq}{ppair}
\Ptwo(t) \approx \paren{h \over \xt} \paren{a^d \over \xt^d} \paren{t
\over \ta} \paren{\Qbare \ta} 
        \approx Q\ta \power{t}{\ta}{1-(d+1)/z}  \period
                                                                \end{eq}
%-----------------------------------------------------------------------
This implies a characteristic timescale, $\tstartwo$, defined such that
$\Ptwo(\tstartwo)\equiv 1$,
%_______________________________________________________________________
                                                \begin{eq}{tstartwo}
\tstartwo = \ta\,\paren{1 \over Q\ta}^{z /(z-d-1)} \period
                                                                \end{eq}
%-----------------------------------------------------------------------
For $t>\tstartwo$ any pair initially within diffusive range will
definitely have reacted; this tells us that kinetics must have 2nd
order DC form for such times. Thus the DC regime of eq.
\eqref{brown-scroll} begins only at $\tstartwo$; for shorter times,
$t<\tstartwo$, since $\Ptwo(t)\ll 1$, correlations are little
disturbed from equilibrium and it follows that MF kinetics apply,
$\ktwo = \Qbare h a^d$.  In fact, for sufficiently small $Q$ (``weak
systems'') $\tstartwo$ will exceed $\tstarmany$ in which case the 2nd
order DC regime will disappear.  In later sections we will carefully
distinguish between this case and the case of ``strong systems''
($\tstartwo<\tstarmany$).

\subsubsection*{Interfacial Kinetics: the Technical Difficulties}

In the present work we will develop a near-exact formalism to
justify these scaling arguments.  The difficulty is the many-body
character of this problem.  Consider for example the reaction rate per
unit area, $\Rtdot$.  This equals the number of reactive A-B pairs per
unit area which are in contact at the interface, $\rhoabs(t)$,
multiplied by the local reactivity $\Qbare$:
%_______________________________________________________________________
                                                \begin{eq}{rate}
{d\Rt \over dt} = \lambda \rhoabs(t) \comma\gap
                 \lambda\equiv  \Qbare h a^d = Q a^{d+1}
                                                                \end{eq}
%-----------------------------------------------------------------------
where the quantity $\lambda$ will turn out to be a natural coupling
constant in our theory.  Now $\rhoabs(t)$ is the two-body density
correlation function $\rhoab(\ra,\rb;t)$ (the number of A-B pairs at
$\ra, \rb$ per unit volume squared) evaluated at the interface,
$\ra=\rb=0$:
%_______________________________________________________________________
                                                \begin{eq}{rhoabs-intro}
\rhoabs(t) \equiv \rhoab(0,0;t) \period
                                                                \end{eq}
%-----------------------------------------------------------------------
We take the origin of our coordinate system to lie on the interface
plane and we have used translational invariance in the directions
parallel to the interface (hence $\rhoabs(t)$ is spatially uniform).
One sees that to determine the reaction rate we need information on
the two-body density correlation function.  However, any dynamical
equation for the latter inevitably involves three-body correlation
functions $\rhoaba,\rhobab$. The dynamics of these objects in turn
involve higher order correlations, and so forth. This hierarchical
structure is the signature of the many-body nature of the problem.

How can a theory deal with these many-body complexities?  One possible
approach \citeben{fredricksonmilner:reactiface_timedept}, a mean field
approximation, would be to assume $\rhoabs(t) = [\ns(t)]^2$, where
$\ns(t)\equiv n(\r=0)$ is the density of A (or B) reactants at the
interface.  This approximation cannot always be valid: for example, in
the compact case, $d+1<z$, this would disagree with the short time 2nd
order DC behavior of eq. \eqref{brown-scroll} since in this regime the
density field is unchanged from equilibrium, $\ns(t)\approx n(0)$;
hence the assumption $\rhoabs(t) = [\ns(t)]^2$ would wrongly yield
$\Rt\twid t$.  Does this approximation make sense at longer times?
Now since we have established (eq. \eqref{glass-in-the-wine}) the asymptotic
law $\Rtdot \twid n t^{(1-z)/z}$, this approximation would then imply
$\ns(t) \sim t^{(1-z)/(2z)}$ which as we will see is correct for the
non-compact case only.  For the compact case, the long time decay of
$\ns(t)$ is in fact determined by the rate at which fluctuations in
the initial distribution of A and B reactants decay.  This gives rise
to a different decay law, invalidating the local mean field
approximation.

To see how densities at the interface, $\ns(t)$, decay for large
times, consider a simple generalization of the argument of Ovchinnikov
and Zeldovich and Toussaint and Wilczek, extended to the interface
problem.  Consider a region at the interface of volume $\Omega$, half
of which is on the A side and half on the B side.  The difference
$\delta N(t) \equiv N_A(t)-N_B(t)$ between the number of A and B in
$\Omega$ is initially of order $\sqrt{n \Omega}$.  Now fluctuations in
$\delta N(t)$ can decay by diffusion only.  Only if $\Omega$ is
smaller than $\xt^d$ did these fluctuations have sufficient time to
have decayed by time $t$.  For bigger regions, $\delta N(t)\approx
\delta N(0) \approx
\sqrt{n \Omega}$.  Thus reactant densities at the interface, for example, 
cannot decay faster than $\sqrt{n \xt^{d}}/ \xt^{d} \twid \sqrt{n}
t^{-d/(2z)}$. In the compact case, $d+1<z$, this is a slower decay
than $[\rhoabs(t)]^{1/2}$.  Thus the local mean field assumption is
wrong, and subtle correlations between reactants determine the long
time decay.  Correspondingly, for the compact case only, there is a
segregation of reactants adjacent to the interface into A-rich and
B-rich regions of size $\xt$.

\ignore{
Many other approximation schemes have been applied in trying to solve
for the reaction kinetics in bulk reaction systems.  
A widely used
approximation is to ignore completely the three-body correlation
functions, reducing thus the many-body problem to a two-body one.
This is a valid approximation in the limit of very dilute reactants
(ref Ben).
}

Various approximation schemes have been used to treat reaction
kinetics in the bulk.  Typically, the three-body density correlation
function is truncated in terms of lower order correlations; this
reduces the hierarchy of reaction-diffusion equations for the
many-body correlation functions to a closed set which are solved
numerically (see ref.\citenum{kotominkuzovkov:book} and references
therein).  The ad-hoc nature of such approximations is balanced by
their success, as judged from direct numerical simulations
\citeben{kotominkuzovkov:book}.  Rigorous  analysis 
was initiated by Doi \citeben{doi:reaction_secondquant1and2} who
developed a general formalism mapping classical many-particle systems
onto quantum field theoretic models.  Doi's formalism has been the
starting point of recent renormalization group approaches to bulk
reacting systems \citeben{ben:aplusa_rapidcomm,lee:aa_rg,leecardy:ab_rg}.

Our approach is rather different to previous ones.  We make a small
number of simple assumptions which on physical grounds we believe are
correct: we assume bounds on certain density correlation functions,
and we assume the reaction rate to be a decreasing function of time
which is asymptotically a power law.  It is possible that these bounds
might be proved rigorously, but we do not attempt this here.  Having
made these assumptions, the subsequent analysis is exact.  In the case
of systems such as reacting polymers which are not point-like (all the
internal polymer degrees of freedom are involved in addition to the
locations of the reactive groups) and for which $z\ne 2$ at small
times, our analysis, though not exact, provides a framework for
calculating all physically interesting quantities.

\vi

The rest of this paper aims to justify the scaling arguments presented
above.  In Section 2 we present an exact mathematical formulation of
the problem.  We significantly simplify the problem in section 3 by
postulating bounds on a three-body density correlation function.
This allows us to solve for the reaction rate.  In Sections 4, 5 and 6
we solve for the reaction rate in the compact, noncompact and marginal
($z=d+1$) cases.  Our results verify the scaling arguments presented
above.  In Sections 7 and 8 the density profile is calculated, including
fluctuation effects and reactant segregation.  We conclude with a
discussion of our results in Section 9.

%************************************************************************************
%************************************************************************************
%************************************************************************************
%************************************************************************************

\section{Interfacial Pair Density, $\rhoabs$}

According to eq. \eqref{rate} the reaction rate is
proportional to the density of A-B pairs which are in contact at the
interface, $\rhoabs(t)$.  In this section, we will obtain an exact
self-consistent integral expression for $\rhoabs$.

We consider the general situation, illustrated in fig. \ref{iface},
where the initial reactant densities $\nainf,\nbinf$ are not
necessarily equal. (The entire discussion of section I treated the
symmetric case $\nainf=\nbinf$ for simplicity.)  Our convention will
always be that $\nbinf\ge \nainf$.  We choose the $d$-dimensional A and
B bulk phases to occupy $x>0$ and $x<0$ respectively, with $x$ being
the direction orthogonal to the interface, and we assume that A and B
species have identical dynamics (\ie dynamical exponent $z$). 

Throughout this paper, we use the convention that superscript $T$
denotes a $d$-dimensional vector lying in the ($d$-1)-dimensional
interface.  Thus by definition the $x$-component of $\rt$ vanishes.

Let us begin by treating the case $z=2$, which is then simply
generalized to arbitrary $z$.  The second-quantization representation
for classical many-particle systems developed by Doi
\citeben{doi:reaction_secondquant1and2} and by 
Zeldovich and Ovchinnikov \citeben{zeldovichovchinnikov:secondquant}
allows us to derive an exact reaction diffusion equation for the
two-body correlation function $\rhoab(\ra,\rb;t)$.  Using Doi's
formalism, we show in Appendix A that for small non-interacting
Fickian molecules ($z=2$) with diffusivity $D$
%_______________________________________________________________________
                                                \begin{eqarray}{rhoab-dot}
\curly{
{\partial \over \partial t}  
- D\,[\nabla_A^2 + \nabla_B^2]
} \rhoab(\ra,\rb;t) &= &
 - \lambda \, \delta (\xa)
\delta (\ra - \rb) \ \rhoab(\ra,\rb;t) \drop
&-&\lambda \, \delta (\xa) \ \rhoabb (\ra,\rb,\ra;t) 
 - \lambda\, \delta (\xb) \ \rhoaba (\ra,\rb,\rb;t) \comma \drop
                                                                \end{eqarray}
%-----------------------------------------------------------------------
with reflecting boundary conditions at $x=0$.  Note the appearance of
the coupling constant $\lambda\equiv \Qbare h a^d$ introduced in eq.
\eqref{rate}.  The 3-body correlation function $\rhoabb
(\ra,\rb,\rbprime;t)$ is the probability density to find an A-B-B
triplet at locations $\ra,\rb,\rbprime$.  A similar definition applies
to $\rhoaba (\ra,\rb,\raprime;t)$.

The sink terms on the right hand side of eq. \eqref{rhoab-dot}
describe the three ways in which reactions can diminish $\rhoab
(\ra,\rb;t)$.  (1) The first {\em two-body} sink term represents
reactions between A-B pairs located at $\ra,\rb$.  The delta functions
restrict reactions to $\ra,\rb$ values such that both A and B are in
contact (\ie within $a$ of one another) and both A and B are within
the interface of width $h$ located at $x=0$.  These restrictions
introduce a factor $h a^d$.  This is a somewhat coarse-grained
description: our ``minimal'' delta-function sinks are appropriate
provided we avoid timescales of order $\th\equiv \ta (h/a)^z$ or
smaller. (2),(3) The remaining two sink terms in eq.
\eqref{rhoab-dot} describe reactions involving just {\em one} particle
of an A-B pair at $\ra,\rb$.  Such a reaction involves a third
particle, weighted by the appropriate 3-body correlation function.
These are {\em many-body} terms; were they absent, one would have a
relatively simple closed 2-body system.  In the next section we will
deal with this difficulty by assuming bounds on the forms of these
3-body correlation functions.

Consider a general value of $z$ now, for which the two particle free
propagator is $\Gt(\ra,\ra';\rb,\rb')$, namely the probability density
an A-B pair is at $\ra,\rb$ at time $t$ given initial location
$\ra',\rb'$, in the {\em absence} of reactions.  Without reactions A
and B particles are statistically independent; thus $\Gt$ can be
written as a product of single particle propagators $\Gtone$\ :
%_______________________________________________________________________
                                                \begin{eq}{green-product}
\Gt(\ra,\ra';\rb,\rb') = \Gtone(\ra,\ra') \ \Gtone(\rb,\rb') \period
                                                                \end{eq}
%-----------------------------------------------------------------------
Since $\Gtone$ has only one characteristic scale, $\xt$, dimensional
analysis dictates the scaling form
%_______________________________________________________________________
                                                \begin{eq}{green}
\Gtone(\r,\r') = \inverse{\xt^d} \ 
g( \r/ \xt ,\,  \r'/ \xt  )
                              \comma \gap 
g(\bu,\bv) \gt 
\casesbracketsii
{f(u_x, v_x)}        {| \bu - \bv | \ll 1 }
{0}                  {| \bu - \bv | \gg 1 }
                                                             \end{eq}
%-----------------------------------------------------------------------
where $f(u_x, v_x)$ is a function of order unity for every value of
its arguments ($u_x$ and $v_x$ are the $x$ components of $\bu$,
$\bv$, respectively).  The fact that $f$ depends on $u_x,v_x$ is a
result of the broken translational invariance in the $x$ direction due
to the reflecting boundary at $x=0$.

Returning to eq. \eqref{rhoab-dot}, we can write a self-consistent
expression for $\rhoab (\ra,\rb;t)$ in terms of the free propagator
$\Gt$.  Setting $\ra=\rb=0$ one obtains
%_______________________________________________________________________
                                                \begin{eq}{exact-rhoabs}
\rhoabs(t) = \nainf \nbinf - \lambda \int_{0}^{t} d\tprime 
\Sd(t-t') \rhoabs(\tprime) - \lambda [\Ima + \Imb] \comma
                                                                \end{eq}
%-----------------------------------------------------------------------
where
%_______________________________________________________________________
                                                \begin{eq}{sd-def}
\Sd(t)  \equiv  \int d\rtprime \Gt(0,\rtprime;0,\rtprime) \approx
{1 \over \xt^{d+1}} 
                                                            \end{eq}
%-----------------------------------------------------------------------
is the two-body ``return probability,'' namely the probability density
an A-B pair is in contact at time $t$ at the interface, given it was
in contact somewhere within the interface at $t=0$.  We have used eq.
\eqref{green} to show that $\Sd(t)\approx 1/\xt^{d+1}$ has the same
scaling form as the return probability in a ($d+1$)-dimensional bulk
problem.  In eq. \eqref{exact-rhoabs}, the two-body integral involving
$\Sd(t)$ represents depletion in the interfacial reactive pair density
$\rhoabs(t)$ due to A-B pairs whose members reacted with one another
at times $t'<t$ and therefore failed to reach the origin at $t$ (see
fig. \ref{sink_terms}).  The terms $\Ima,\Imb$ measure
depletion due to many-body effects: 
%_______________________________________________________________________
                                                \begin{eqarray}{im-def}
\Ima &\equiv& \int_0^t d\tprime \int d\ratprime d\rbprime \Gttprime
(0,\ratprime;0,\rbprime) \rhoabb(\ratprime,\rbprime,\ratprime;\tprime)
\comma \drop
\Imb &\equiv& \int_0^t d\tprime \int d\raprime d\rbtprime \Gttprime
(0,\raprime;0,\rbtprime) \rhoaba(\raprime,\rbtprime,\rbtprime;\tprime) \period
                                                                \end{eqarray}
%-----------------------------------------------------------------------
These integrals subtract off any A-B pair only one member of which was
involved in a reaction an earlier time (see fig.
\ref{sink_terms}).

We will see later that the two-body integral involving $\Sd(t)$ in eq.
\eqref{exact-rhoabs} is important at short times; for such times
reaction kinetics are hence like those in a ($d+1$)-dimensional bulk
reaction problem.  At longer times the many-body terms $\Ima,
\Imb$ are always dominant and kinetics cross over to one-dimensional form.

Eqs. \eqref{exact-rhoabs}, \eqref{sd-def} and \eqref{im-def} are
immediately generalized to arbitrary dynamics with arbitrary values of
$z$: one simply replaces the Gaussian ($z=2$) propagator $\Gt$,
describing Fickian diffusion, with the appropriate propagator
describing the dynamics.  Now this would be a true statement for the
abstract concept of small (\ie point-like) molecules obeying $x_t\twid
t^{1/z}$ with $z\ne2$.  However, in practice non-Fickian diffusion
normally results from the small reactive species belonging to a large
structure with complex internal dynamics.  The most important case is
when the reactive group is a single monomer unit belonging to a
polymer chain of N units.  In these cases the dynamics of eq.
\eqref{exact-rhoabs} are not exact because they incorrectly presuppose a
closed relationship in terms of the degrees of freedom specifying the
location of the reactive species only.  A proper treatment must first
average out the other degrees of freedom (e.g. the locations of the
other $N-1$ monomers in the polymer case); this is non-trivial and
requires renormalization group (RG) methods
\citeben{ben:cycprl}.  However, RG studies of 2-body bulk polymer
reaction kinetics \citeben{ben:cycprl,ben:interdil} indicate that the
basic physics is completely captured by the approximate closing of the
system in terms of these coordinates only: correct scaling behaviors
are obtained, only the prefactors being unreliable.  These issues are
discussed in detail in refs.
\citenum{ben:meltsinternaljcp,ben:cycprl,ben:interdil}.  Therefore for
the remainder of this paper we assume the validity of eqs.
\eqref{exact-rhoabs}, \eqref{sd-def} and \eqref{im-def} for any value
of $z$.

%************************************************************************************
%************************************************************************************
%************************************************************************************
%************************************************************************************

\section{Structure of Many-Body Integral Terms $\Ima,\Imb$}

In Section 2 we derived a self-consistent solution for the interfacial
reactive pair density $\rhoabs(t)$, eq.
\eqref{exact-rhoabs}.  Unfortunately this is not in a closed form for
$\rhoabs(t)$, since the many-body terms involve higher order
correlation functions.  In this section we introduce our three simple,
physically motivated assumptions.  These enable us to express
$\Ima,\Imb$ in terms of $\rhoabs(t)$, which in the following section
will allow us to obtain a closed solution for $\rhoabs$.  Most
calculational details will be left for Appendix B.

Consider the three-body correlation function $\rhoabb$ appearing in
$\Ima$ of eq. \eqref{im-def}.  Let us introduce 
the conditional density of B particles at
$\rb$ given an A-B pair at the origin, 
%_______________________________________________________________________
                                                \begin{eq}{pea}
\rhobab(\rb | 0,0;t) \equiv {\rhoabb(0,\rb,0;t) \over \rhoabs(t)} \period
                                                                \end{eq}
%-----------------------------------------------------------------------
Noting that translational invariance parallel to the interface plane
allows the replacement $\rhoabb(\ratprime,\rb',\ratprime;\tprime)\gt
\rhoabb(0,\rbprime - \ratprime,0;\tprime)$, we can express $\Ima$ as
%_______________________________________________________________________
                                                \begin{eqarray}{broom}
\Ima = \int_0^t d\tprime \int d\rbprime \ F_{t-t'}(\rbprime)
\ \rhobab(\rbprime | 0,0;t')\ \rhoabs(t') \comma
                                                \drop
F_{t-t'} (\rbprime ) \equiv  \int d\ratprime\  G_{t-t'}
(0,\ratprime;0,\rbprime+\ratprime) \period
                                                                \end{eqarray}
%-----------------------------------------------------------------------

In fig. \ref{world_lines} we identify two physically distinct
space-time regions which contribute to $\Ima$ in the
$\rbprime,\tprime$ integration of eq. \eqref{broom}.  The assumptions
we are about to introduce are based on the following expectations
about the behavior of the conditional 3-body density in these two
regions.  In region I, defined by points with $x$-coordinate
$\xb'>x_{\tprime}$, the conditional density at time $\tprime$
approximates its far field value, $\rhobab(\rbprime | 0,0;\tprime)
\approx \nbinf$.  This is because far into region I such locations
$\rb'$ are beyond diffusional range of the interface: hence density
correlations at $\rb'$ cannot have been influenced by reaction events
during $(0,t')$.  On the other hand, in region II ($\xb' <
x_{\tprime}$), this conditional density will be strongly influenced by
such reaction events.  Whatever this density field may be, we expect
that its maximum will never be greater than a value of the order of
$\nbinf$.  Reactions tend to reduce densities, but we do not exclude
the possibility that subtle B-A-B correlations could locally elevate
the field somewhat.

Let us now translate the above general physical expectations into two
specific assumptions on the conditional density field.  Simultaneously
we introduce a third assumption, concerning $\rhoabs(t)$.

{\bf Assumption 1.} There exists a positive finite constant $U$, such that:
%_______________________________________________________________________
                                                \begin{eq}{ass-one}
{\rhobab(\rbprime | 0,0;t') \over \nbinf} \leq U \period
                                                                \end{eq}
%-----------------------------------------------------------------------
This amounts to assuming that irrespective of what reaction-induced
correlations exist between points $0$ and $\rbprime$, the conditional
density of B particles at $\rbprime$ will always be less than, or at
most of the order of, the far-field density of B reactants in the B
bulk.\\

{\bf Assumption 2.} There exists a positive finite constant $L$, such that:
%_______________________________________________________________________
                                                \begin{eq}{ass-two}
{\rhobab(\rbprime | 0,0;t') \over \nbinf} \geq L \comma \gap {\rm for
\ }\, {\xb' \over x_{\tprime}} > 1 \period
                                                                \end{eq}
%-----------------------------------------------------------------------
Roughly speaking this amounts to assuming that points in region I are
uncorrelated with the interface.\\

{\bf Assumption 3.} $\rhoabs(t)$ is a decreasing function of time which
is asymptotically a power law. \\

Assumptions 1 and 2 immediately imply the same two assumptions but
with A and B interchanged (since A and B are arbitrarily chosen labels).
That is, the field $\rhoaba(\raprime | 0,0;t')$ appearing in $\Imb$
obeys two assumptions analogous to 1 and 2.

Based on these assumptions, we show in Appendix B that the
contribution to $\Ima$ from integration over region I is a fraction of
order unity of the value of $\Ima$.  (Equivalently, a fraction of
order unity of the A-B interface pairs which involve one previously
reacted A and are subtracted off by $\lambda \Ima$, involve a B member
originating from region I.)  Therefore, since assumptions 1 and 2
imply that $\rhobab$ equals $\nbinf$ in region I to within a finite
prefactor bounded above and below, it follows that if we replace
$\rhobab(\rbprime | 0,0;\tprime) \gt \nbinf$ in the integrand of
$\Ima$ in eq. \eqref{broom}, the result will equal the actual value of
$\Ima$ to within a (time-dependent) prefactor of order unity,
$\alpha(t)$.  Making this replacement, using eq.
\eqref{green-product}, and performing an analogous replacement for $\Imb$, one
obtains
%_______________________________________________________________________
                                                \begin{eq}{simple-Ima}
\Ima = \alpha(t)\ \nbinf \int_0^t d\tprime \Sone(t -\tprime) \rhoabs(\tprime)
        \comma \gap
\Imb = \beta(t)\  \nainf \int_0^t d\tprime \Sone(t -\tprime) \rhoabs(\tprime)
\comma
                                                                \end{eq}
%-----------------------------------------------------------------------
where $\alpha(t), \beta(t)$, are bounded positive functions of order
unity,
%_______________________________________________________________________
                                                \begin{eq}{alpha-bounds}
\alpha_{\rm min} \leq \alpha(t) \leq \alpha_{\rm max} \comma \gap
\beta_{\rm min}
\leq \beta(t) \leq \beta_{\rm max} \period
                                                                \end{eq}
%-----------------------------------------------------------------------
Here $\alpha_{\rm min}, \alpha_{\rm max}, \beta_{\rm min}, \beta_{\rm
max}$ are finite positive constants.  The one-dimensional return
probability $\Sone(t)$ is defined as
%_______________________________________________________________________
                                                \begin{eq}{sone}
\Sone (t)  \equiv \int d\ratprime \Gt^{(1)}(0,\ratprime) 
           \approx {1 \over \xt} \period 
                                                                \end{eq}
%-----------------------------------------------------------------------
It measures the probability a reactant initially at the interface
returns to the interface after time $t$.  The scaling form,
$\Sone\twid 1/\xt$, is easily derived from the scaling form of the
propagator $\Gt^{(1)}$, eq. \eqref{green}.

%************************************************************************************
%************************************************************************************
%************************************************************************************
%************************************************************************************

\section{Reaction Rate in Compact Case ($d+1<z$)}

Having expressed the many-body terms $\Ima,\Imb$ in terms of the
interfacial reactive pair density $\rhoabs(t)$, we can solve for the
reaction rate per unit area $\Rdot = \lambda
\rhoabs(t)$.  According to the results of the previous section, eq.
\eqref{simple-Ima}, the self-consistent solution for $\rhoabs(t)$, eq.
\eqref{exact-rhoabs}, can be written
%_______________________________________________________________________
                                                \begin{eqarray}{simple-rhoabs}
\rhoabs(t) &=& \nainf \nbinf - \lambda \int_0^t d\tprime \Sd(t-\tprime)
\rhoabs(\tprime) - \lambda n(t) \int_0^t
d\tprime \Sone(t -\tprime) \rhoabs(\tprime) \comma \gap \ddrop
n(t) &\equiv& \alpha(t) \nbinf + \beta(t) \nainf \period
                                                      \end{eqarray}
%-----------------------------------------------------------------------
This ``solution'' of course involves the unknown function $n(t)$.
From the arguments of the previous section following from our
assumptions 1 and 2, we know that $n(t)$ is bounded above and below.
Now according to assumption 3, asymptotically $\rhoabs(t) \sim
t^{-\delta}$ with $\delta > 0$.  Substituting this power law in eq.
\eqref{simple-rhoabs}, and substituting $\xt \twid t^{1/z}$ in
the scaling forms of $\Sd(t)$ and $\Sone(t)$ from eqs. \eqref{sd-def}
and \eqref{sone}, one finds that as $t\gt \infty$ the many-body term
dominates over the other time-dependent terms in eq.
\eqref{simple-rhoabs}, and up to a constant prefactor is equal to
$n(t) t^{1-1/z-\delta}$.  It follows that at long enough times the
many-body term must equal the first term on the rhs of eq.
\eqref{simple-rhoabs}, $\nainf \nbinf$, plus higher order corrections.
Since $n(t)$ is bounded, this implies that $\delta =1-1/z$ and that
$n(t)$ tends to a constant at long times, $n(\infty)$.  We will prove
in section 7 that this constant is none other than the reactant
density in the more {\em dense} of the two phases:
%_______________________________________________________________________
                                                \begin{eq}{n}
\nofinfinity = \nbinf \period
                                                                \end{eq}
%-----------------------------------------------------------------------
We remind the reader of our convention throughout this study,
$\nbinf\ge \nainf$.

Laplace transforming eq. \eqref{simple-rhoabs}, $t \rightarrow E$, and
recalling that $\Rdot = \lambda \rhoabs(t)$, it is simple to obtain
the following self-consistent relation for the Laplace transform of
the reaction rate per unit area, $\Rdot(E)$:
%_______________________________________________________________________
                                                \begin{eq}{pierre}
\Rdot(E) = {\lambda \nainf \nbinf \over E \square{1+ \lambda \Sd(E)
+\lambda \nbinf \gamma(E) \Sone(E) }}   
 \comma \ \ \ \ 
\gamma(E) \equiv {n(E) {\rm *} [\Sone(E) \Rdot(E)] \over \nbinf \Sone(E) \Rdot(E)}
 \period
                                                                \end{eq}
%-----------------------------------------------------------------------
Here, $*$ indicates convolution in Laplace space.  The function
$\gamma(E)$ has a simple form for small $E$; since, by virtue of eq.
\eqref{n}, $n(E \rightarrow 0) = \nbinf / E$, then from eq.
\eqref{pierre} one has
%_______________________________________________________________________
                                                \begin{eq}{gamma}
\gamma(E) = 1 + O(E)  \gap (E \rightarrow 0) \period
                                                                \end{eq}
%-----------------------------------------------------------------------
Now from eqs. \eqref{xt}, \eqref{sd-def} and \eqref{sone}, $\Sd(t)$ and
$\Sone(t)$ are algebraic in time.  Their Laplace transforms have the
form $\Sd(E)\twid E^{(d+1)/z - 1}$ (valid only in compact dimensions,
$d+1<z$) and $\Sone(E)\twid E^{1/z - 1}$ (always valid).  This section
is concerned with the compact case; then we can rewrite $\Rdot(E)$ in
two ways:
%_______________________________________________________________________
                                         \begin{eqarray}{food}
\Rdot(E) &\approx& {\lambda \nainf \nbinf \over E \curly{1+ \paren{E
                        \tstartwo}^{[(d+1)/z]-1} + \gamma(E) \paren{E
                                                \tstarmany}^{(1/z)-1}}} \drop 
         &\approx& {\lambda \nainf \nbinf \over E
                        \curly{1+\paren{E\tstartwo}^{[(d+1)/z]-1}
                        \square{1+ \gamma(E)\paren{E\tl}^{-d/z}}}}      
           \comma
                                                                \end{eqarray}
%-----------------------------------------------------------------------
where 
%_______________________________________________________________________
                                                \begin{eq}{clock}
\tl \equiv \ta\, \paren{1 \over \nbinf a^d}^{z/d}
                         \comma \gap
\tstarmany \equiv \ta\, \paren{1 \over Q \ta \, \nbinf a^{d}}^{z/(z-1)}
                         \comma \gap
\tstartwo \equiv \ta\, \paren{1 \over Q \ta}^{z/[z-(d+1)]}   \comma
                                                                \end{eq}
%-----------------------------------------------------------------------
are essentially the three naturally occurring timescales introduced in
section 1, generalized to the case of unequal initial reactant
densities ($\nbinf \ge \nainf$).  It is important to note that the
characteristic density determining these timescales is that of the
{\em denser} bulk phase B.  In eq. \eqref{food} for simplicity we have
neglected numerical prefactors in the terms in the denominator.

Note that the three characteristic timescales obey
%_______________________________________________________________________
                                                \begin{eq}{mouse}
\tstarmany = {\paren{\tstartwo}}^{1-d/(z-1)}
                {\paren{\tl}}^{d/(z-1)} 
                                                                \end{eq}
%-----------------------------------------------------------------------
which implies that the magnitude of $\tstarmany$ always lies between
those of $\tstartwo$ and $\tl$.  Hence there are only 2 cases (see
fig. \ref{fund_phase}).  (a)
For strongly reactive (``strong'') systems, $Q>\Qstar$, the ordering
of timescales is $\tstartwo < \tstarmany < \tl$. (b) For ``weak''
systems, $Q<\Qstar$, one has $\tl< \tstarmany < \tstartwo$.  The
boundary between strong and weak regimes is defined by a critical
effective local reactivity, at which $\tstartwo = \tstarmany = \tl$\ :
%_______________________________________________________________________
                                                \begin{eq}{lstar}
\Qstar \ta \equiv (\nbinf a^d)^{[z-(d+1)]/d} \period
                                                                \end{eq}
%-----------------------------------------------------------------------

{\bf (a) Strong Systems} : $Q>\Qstar$, $\tstartwo < \tstarmany < \tl$.  Before
evaluating $\Rtdot$ in different time regimes, we note that the many
body term (the third term in the denominator in eq. \eqref{food}) is
unimportant whenever $E\tl\gg 1$ (corresponding to $t \ll \tl$).
Consider the term (that involving $\Sone$) in eq.
\eqref{simple-rhoabs} from which this many body contribution is
derived.  Now imagine replacing $n(t)$ in this term by its maximum
value, $n(t)
\gt n_{\rm max} \equiv \alpha_{\rm max} \nbinf + \beta_{\rm max}
\nainf$, such that $n(E) = n_{\rm max}/E$ and hence $\gamma(E) =
n_{\rm max} / \nofinfinity \approx 1$.  It would then indeed follow
that the many body term in eq. \eqref{food} is higher order for $E\tl\gg 1$. 
Clearly, then, this must always be true.

Consider firstly short times, $E^{-1} \ll \tl$.  The many body term
can then be neglected.  Considering the two cases $E^{-1}\ll
\tstartwo$ and $\tstartwo \ll E^{-1} \ll \tl$, respectively, inverse
Laplace transformation of eq. \eqref{food} yields
%_______________________________________________________________________
                                                \begin{eq}{pencil}
\Rtdot = \ksecond \nainf \nbinf 
                        \comma \gap
 \ksecond \approx \casesbracketsii{\lambda}{t \ll \tstartwo}
                  {d\xt^{d+1}/dt \twid t^{(d+1)/z-1}}{\tstartwo \ll t \ll \tl} 
                         \period
                                                                \end{eq}
%-----------------------------------------------------------------------
These are 2nd order rate kinetics.  An initial MF regime is followed
at $\tstartwo$ by a DC regime.  Notice that the timescale $\tstarmany$
is irrelevant.  We remind the reader that our analysis does not
describe times less than $\th$ (see comments following eq.
\eqref{rhoab-dot}); hence, if $Q$ is so great that
eq. \eqref{clock} implies $\tstartwo<\th$, then eq. \eqref{pencil}
correctly describes the second DC regime only.

Now consider very long times $E^{-1} \gg \tl$; the many-body term in
eq. \eqref{food} is then dominant.  Since at long enough times we may
replace $\gamma(E)\gt 1$ as discussed, one now finds {\em first} order
kinetics:
%_______________________________________________________________________
                                                \begin{eq}{first-order-general}
\Rdot = \kfirst \nainf \comma  \gap
           \kfirst \approx {d\xt \over dt} \twid t^{1/z-1}
                                        \gap (t\gg \tl)\period  
                                                                \end{eq}
%-----------------------------------------------------------------------
Thus, at long enough times the reaction rate is controlled by the
diffusion to the interface of the more {\em dilute} A species.

{\bf (b) Weak Systems}: $Q<\Qstar$, $\tl< \tstarmany < \tstartwo$.
Now the many-body term in the curly brackets in eq. \eqref{food} is
much smaller than 1 whenever $E^{-1} \ll \tstarmany$ (this can be seen
by replacing $n(t)$ with its maximum value as we did for (a) above).
But for such $E$ values, it is automatically true that $E\tstartwo\ll
1$ by virtue of the definition of weak systems
($\tstartwo>\tstarmany$), and hence the 2-body term in eq.
\eqref{food} is also much smaller than unity.  It follows that MF 2nd
order kinetics pertain for all times less than $\tstarmany$
%_______________________________________________________________________
                                                \begin{eq}{sugar}
\Rtdot = \ksecond \nainf \nbinf \comma \gap
        \ksecond \approx \lambda  \gap  (t \ll \tstarmany) \period
                                                                \end{eq}
%-----------------------------------------------------------------------

Notice that the 2-body and many-body terms are both proportional to
negative powers of $E$, and that the magnitude of the many-body term's
exponent is the greatest of the two.  Now consider $E^{-1}\gg
\tstarmany$, when the many body term is much bigger than unity.  It
follows that this term is then also much bigger than the 2-body term,
because $\tstarmany<\tstartwo$ for these weak cases.  Thus kinetics
are first-order for $t\gg \tstarmany$:
%_______________________________________________________________________
                                                \begin{eq}{skim-milk}
\Rtdot = \kfirst \nainf \comma \gap 
   \kfirst \approx {d\xt \over dt} \twid t^{1/z -1}
                 \gap (t \gg \tstarmany) \period
                                                                \end{eq}
%-----------------------------------------------------------------------
To obtain eq. \eqref{skim-milk} we have replaced $\gamma(E)\gt 1$ for
small $E$.  For weak systems, neither $\tstartwo$ nor $\tl$ are
relevant.  The reactivity is so small that the two-body term is never
relevant, and 2nd order DC kinetics are absent.

%***********************************************************************************

%***********************************************************************************
%***********************************************************************************
%***********************************************************************************
%~~~~~~~~~~~~~~~~~~~~~~~~~~~~~~~~~~~~~~~~~~~~~~~~~~~~~~~~~~~~~~~~~~~~~~~~~~~~~~~~~~~~
%~~~~~~~~~~~~~~~~~~~~~~~~~MARGINAL CASE~~~~~~~~~~~~~~~~~~~~~~~~~~~~~~~~~~~~~~~~~~~~~~
%***********************************************************************************
%***********************************************************************************

\section{Reaction Rate in Marginal Case ($d+1=z$)}

The previous section dealt with low compact dimensions, for which the
2-body return probability in Laplace space had the form $\Sd(E)\twid
E^{(d+1)/z - 1}$.  For $d+1\ge z$, this is no longer true.  In this
section we consider the marginal case, $d+1=z$; thus $\Sd\approx
1/\xt^{d+1}\twid 1/t$, giving
%_______________________________________________________________________
                                                \begin{eq}{cook}
\Sd(E) \approx \int_{\th}^\infty \, dt \, e^{-E t} {\th \over h^{d+1}\ t}
   \approx {\th \over h^{d+1} } \ \ln [1/E\th] \gap (E\th \ll 1) \period
                                                                \end{eq}
%-----------------------------------------------------------------------
We have introduced a cut-off at $t=\th$; at shorter times $\Sd(t)$
crosses over to a form appropriate to a $d$-dimensional bulk problem,
$\Sd \approx 1/(h\xt^d)$ whose time integral gives a contribution of
the same order as that from the lower limit in eq. \eqref{cook}.  

Aside from this modification, all steps leading to eq. \eqref{food} of
the compact case are unchanged: the expression for the reaction rate
$\Rdot(E)$ (eq. \eqref{pierre}) remains valid, and the form of
$\Sone(E)$ is unchanged.  Thus,
%_______________________________________________________________________
                                                \begin{eqarray}{terrible-confusion}
\Rtdot(E) &\approx& {\lambda \nainf \nbinf \over E \curly{
1+ {\ln(1/E\th) \over
                   \ln(e \tstartwo/\th)} + \gamma(E) (E \tstarmany)^{1/z-1}}
} \drop
     &\approx&  {\lambda \nainf \nbinf \over
                  E\curly{1+ {\ln(1/E\th) \over
                \ln(e\tstartwo/\th)}\square{
1+\gamma(E){(E\tl)^{-d/z} \over \ln(1/E\th)} 
} } }  \comma
                                                                \end{eqarray}
%-----------------------------------------------------------------------
Here the definitions of $\tl$ and $\tstarmany$ are unchanged from the
compact case (eq. \eqref{clock}), but now
%_______________________________________________________________________
                                                \begin{eq}{tstartwo-marginal}
\tstartwo \equiv (\th/e)\  e^{1/(Q \ta)} \period
                                                                \end{eq}
%-----------------------------------------------------------------------

Let us define $\Tl$ to be the time such that for $E<\Tl^{-1}$ the
many-body term ($\propto \Sone(E)$) dominates over the two-body term
($\propto \Sd(E)$) in the denominator of eq.
\eqref{terrible-confusion}\ :
%_______________________________________________________________________
                                                \begin{eq}{Tl}
{\Tl \over \tl} = \square{\, \ln (e \Tl / \th) \,}^{z/d} \period
                                                                \end{eq}
%-----------------------------------------------------------------------
(We have included factors of $e$ in the definitions of $\tstartwo$ and
$\Tl$ above simply to ensure continuity of reaction rates; see eqs.
\eqref{croissant} and \eqref{pepsi} below.)

Analogously to the compact case, the condition $\tstartwo = \tstarmany
= \Tl$ defines a critical reactive strength $\Qstar$,
%_______________________________________________________________________
                                                \begin{eq}{Qstar-marginal}
\Qstar \ta \equiv {1 \over \ln \, [e \Tl / \th]} \comma
                                                                \end{eq}
%-----------------------------------------------------------------------
defining the boundary between ``weak'' and ``strong'' kinetics (see
fig. \ref{fund_phase}), for
which it can be shown that the 3 relevant timescales have the same
orderings as for the compact case.

{\bf (a) Strong Systems : $Q > \Qstar, \ \tstartwo < \tstarmany <
\Tl$.} Consider first short times, $E^{-1} \ll \Tl$.  Similar
reasoning as for the compact cases implies that the many-body term in
eq. \eqref{terrible-confusion} can be neglected for such $E$ values.
Considering the two cases $E^{-1} \ll \tstartwo$ and $E^{-1} \gg
\tstartwo$ one obtains
%_______________________________________________________________________
                                                \begin{eq}{croissant}
\Rtdot = \ksecond \nainf \nbinf \comma \gap \ksecond \approx
\casesbracketsii{\lambda}{t \ll \tstartwo}
                {h^{d+1}/[\,\th \ln(e t/\th)\,]}
                                {\tstartwo \ll t \ll \Tl} 
                                                              \end{eq}
%-----------------------------------------------------------------------
The logarithm arises after inverse Laplace transformation of $1/\{E
\ln(1/E\th)\}$ which gives $1/\ln (t/\th)$ for $t\gg \th$.  This is
shown in appendix C.

For long times, $E^{-1} \gg \Tl$, the many-body term dominates.  Using
$\gamma(E)\approx 1$ for small enough $E$, which is easily
demonstrated using similar arguments to those for the compact case,
one finds first-order DC kinetics which are no different in structure
to those for the compact case (see eqs. \eqref{first-order-general}
and \eqref{skim-milk}):
%_______________________________________________________________________
                                                \begin{eq}{pepsi}
\Rtdot = \kfirst \nainf \comma \gap \kfirst \approx {d \xt \over dt}
\twid t ^{1/z-1}  \gap (t \gg \Tl) \period
                                                                \end{eq}
%-----------------------------------------------------------------------

{\bf (b) Weak systems : $Q < \Qstar, \ \Tl < \tstarmany <
\tstartwo$}.  For small times, $E^{-1} \ll \tstarmany$, the many body term
is much less than unity; this is also true of the 2-body term since
$\tstartwo>\tstarmany$ (definition of weak system).  On the other hand,
when $E^{-1} \gg \tstarmany$, the many body term is much larger than
unity; it is also much bigger than the logarithmic 2-body term since
$E^{-1} \gg \Tl$ follows automatically, because $\Tl<\tstarmany$.
Thus
%_______________________________________________________________________
                                           \begin{eq}{cocoa}
\Rtdot = \casesbracketsii{\ksecond \nainf \nbinf \comma \gap \ksecond
                                   \approx \lambda}{t \ll \tstarmany}
                         {\kfirst \nainf \comma \gap \kfirst \approx
                               d\xt / dt \twid t^{1/z-1}}{t \gg \tstarmany}
                                         \period
                                                                \end{eq}
%-----------------------------------------------------------------------

%***********************************************************************************

%***********************************************************************************
%***********************************************************************************
%***********************************************************************************
%***********************************************************************************
%***********************************************************************************

\section{Reaction Rate in Noncompact Case ($d+1>z$)}

In this section high non-compact dimensions are considered, $d+1>z$.
Perhaps the commonest physical example of small molecules ($z=2, d=3$)
belongs to this class.  Mathematically, the only distinguishing
feature is that the Laplace transform of the 2-body return probability
is now dominated by small times, since $\Sd(t)$ of eq. \eqref{sd-def}
now decays faster than $1/t$ for times $t>\th$:
%_______________________________________________________________________
                                           \begin{eq}{salt-on-food}
\Sd(E) \approx \int_{\th}^\infty \, dt \, e^{-E t} 
                       \inverse{h^{d+1}}\ \power{\th}{t}{(d+1)/z}
                   \approx {\th \over h^{d+1} } 
                        \gap (E\th \ll 1, \ z<d+1<z+1) \period
                                                                \end{eq}
%-----------------------------------------------------------------------
The above result is determined by the dominant cut-off at $t=\th$.  In
fact it is valid only provided $d<z$ because only then is the $t<\th$
time integral dominated by its {\em upper} limit, $t=\th$: for these
smallest times (which have been neglected in the original statement of
our model, eq. \eqref{rhoab-dot}) one has in effect an infinite {\em
bulk} reaction problem.  It is as if the interface were infinitely
large.  Correspondingly, the true return probability is $\Sd\approx
1/(h\xt^d)$ for $t<\th$.  When time integrated, for dimensions so high
that even bulk reaction kinetics are non-compact, $d>z$, the lower
cut-off at $\ta$ is now dominant, $\int_{\ta}^{\th} e^{-Et} \Sd dt
\approx \ta \Sd(\ta)$ for $E\th\ll 1$.  This contribution now exceeds
that displayed in eq. \eqref{salt-on-food}, and one has
%_______________________________________________________________________
                                                \begin{eq}{pepper-on-toast}
\Sd(E) \approx {\ta \over h a^d } 
                        \gap (E\th \ll 1, d>z) \period
                                                                \end{eq}
%-----------------------------------------------------------------------
 
Consider firstly $z<d+1<z+1$.  The reaction rate in Laplace space of
eq. \eqref{pierre} now reads
%_______________________________________________________________________
                                                \begin{eq}{drink}
\Rtdot(E) \approx {
\lambda \nainf \nbinf
                 \over  
E\curly{
1+ \Qbare \ta (a^d\th/h^d\ta) + \gamma(E) (E \tstarmany)^{1/z-1} 
}
}
                 \period
                                                             \end{eq}
%-----------------------------------------------------------------------
Here $\gamma(E)$ is the quantity defined in eq. \eqref{pierre} and, as
for the compact and marginal cases, it can be shown that
$\gamma(E)\approx 1$ for small enough $E$.  Thus for $\Qbare \ta <
(a/h)^{z-d}$

%*********OLD LONG FORMAT*****************************************************
\ignore{
%_______________________________________________________________________
                                                \begin{eq}{noncompact-results}
\Rtdot = \casesbracketsii{\ksecond \nainf \nbinf \comma \ \  \ksecond
\approx \lambda}{t \ll \tstarmany} 
{\kfirst \nainf \comma \ \  \kfirst \approx d\xt / dt \twid t^{1/z-1}}
{t \gg \tstarmany}\  \
\paren{\Qbare \ta < (a/h)^{z-d} \comma z<d+1<z+1}                  
                                              \end{eq}
%-----------------------------------------------------------------------
whilst for $\Qbare \ta > (a/h)^{z-d}$ one has
%_______________________________________________________________________
                                                \begin{eqarray}{noncompact-results-2}
\Rtdot && = \casesbracketsii{\ksecond \nainf \nbinf \comma \gap  \ksecond
\approx h^{d+1}/\th}{t \ll \th (\nbinf h^d)^{z/(1-z)}} 
{\kfirst \nainf \comma \gap \kfirst \approx d\xt / dt \twid
t^{1/z-1} \ \ }
{t \gg \th (\nbinf h^d)^{z/(1-z)}} 
  \ddrop
&&\gggap\ggap \ \ \ 
\paren{\Qbare \ta > (a/h)^{z-d} \comma z<d+1<z+1}                  
                                                                \end{eqarray}
%-----------------------------------------------------------------------

Now consider the highest dimensions, $d>z$.  Then $\Rtdot(E)$ is as in
eq. \eqref{drink}, except one replaces $\Qbare \ta
(a^d\th/h^d\ta)\gt\Qbare\ta$, leading to 
%_______________________________________________________________________
                                                \begin{eq}{dirty-table}
\Rtdot = \casesbracketsii{\ksecond \nainf \nbinf \comma \ \  \ksecond
\approx \lambda}{t \ll \tstarmany} 
{\kfirst \nainf \comma \ \  \kfirst \approx d\xt / dt \twid t^{1/z-1}}
{t \gg \tstarmany}\  \
\paren{d>z}                  \period
                                              \end{eq}
%-----------------------------------------------------------------------
} %end \ignore

%*********NEW FORMAT*****************************************************

%_______________________________________________________________________
                                                \begin{eq}{noncompact-results}
\Rtdot = \casesbracketsii{\ksecond \nainf \nbinf \comma \ \  \ksecond
\approx \lambda}{t \ll \tstarmany} 
{\kfirst \nainf \comma \ \  \kfirst \approx d\xt / dt \twid t^{1/z-1}}
{t \gg \tstarmany}\ 
                                              \end{eq}
%-----------------------------------------------------------------------
whilst for $\Qbare \ta > (a/h)^{z-d}$ one has
%_______________________________________________________________________
                                                \begin{eq}{noncompact-results-2}
\Rtdot  = \casesbracketsii{\ksecond \nainf \nbinf \comma \gap  \ksecond
\approx h^{d+1}/\th}{t \ll \th (\nbinf h^d)^{z/(1-z)}} 
{\kfirst \nainf \comma \gap \kfirst \approx d\xt / dt \twid
t^{1/z-1} \ \ }
{t \gg \th (\nbinf h^d)^{z/(1-z)}} 
  \ddrop
                                                                \end{eq}
%-----------------------------------------------------------------------

Now consider the highest dimensions, $d>z$.  Then $\Rtdot(E)$ is as in
eq. \eqref{drink}, except one replaces $\Qbare \ta
(a^d\th/h^d\ta)\gt\Qbare\ta$.  This leads to eq.
\eqref{noncompact-results} which is now valid for all $\Qbare$ values.

%************************************************************************************
%************************************************************************************
%************************************************************************************
%************************************************************************************

\ignore{

This will now be demonstrated explicitly.  We
begin by determining the asymptotic values of the density fields at
the interface, $\nas(\infty)$ and $\nbs(\infty)$.

In section 4 we claimed that the quantity $n(t)$ tends to $\nbinf$ at long
times; this is equivalent to the statement $\gamma(0)=1$ (see eq.
\eqref{pierre}.)  This is the point where we will actually
prove that this is correct.  We will do that by determining the asymptotic
value of the interfacial reactant density $\nas(t) \equiv \na(0;t)$; the
value of $\gamma(0)$ will then immediately follow from eq. \eqref{donut}.

Now, even in the symmetric case $(\nainf=\nbinf)$, the reaction rate becomes
diffusion-controlled at long enough times: every particle that reaches the
interface eventually reacts. This strongly suggests that $\nas$ tends
asymptotically to zero.  However, the way one establishes this fact
mathematically is far less obvious.  We will  show that $\nas(\infty)
= 0 $ by relating $\nas$ to the {\em similar} particle correlation
functions $\rhoaas(t) \equiv \rhoaa(0,0;t)$ and $\rhobbs(t) \equiv
\rhobb(0,0;t)$ through a simple physically motivated assumption which we
present in the following paragraph.

}

\section{Density profile}

We have seen that short time 2nd order reaction kinetics cross over at
a regime-dependent timescale to 1st order diffusion-controlled
kinetics.  This suggests that the density fields on either side of the
interface, $\na(\ra;t) ,\nb(\rb;t)$, are uniform for shorter times but
develop depletion holes at the interface of size $\xt$ when the 1st
order kinetics onset.  To demonstrate this explicitly, we begin by
using Doi's formalism in appendix A to derive the density field
dynamics for small molecules ($z=2)$.  We will then generalize results
to arbitrary dynamical exponent $z$.  For $z=2$, we find
%_______________________________________________________________________
            \begin{eq}{na-nb}
\curly{
{\partial \over \partial t}  
- D\,\nabla_A^2 
}
\na(\ra ;t)  =  - \lambda \delta (\xa) \rhoabs( t )  \comma
\ \ \ \ \
\curly{
{\partial \over \partial t}  
- D\,\nabla_B^2 
}
\nb(\rb ;t) =  - \lambda \delta (\xb) \rhoabs(t)  \period
            \end{eq}
%-----------------------------------------------------------------------
The sink terms on the right hand sides of eq. \eqref{na-nb} are
proportional to the number of $A-B$ pairs which are in contact at the
interface, per unit area.  Noting that translational invariance
parallel to the interface plane implies $\na,\nb$ depend on $\xa$ and
$\xb$ only, eq. \eqref{na-nb} has solution
%_______________________________________________________________________
                                                \begin{eq}{integral-na-nb}
\na(\xa;t) = \nainf - \lambda \int_0^t dt' \Gttprime^{(1)} (\xa) \rhoabs(t')
\comma \ \ \ \ \
\nb(\xb;t) = \nbinf - \lambda \int_0^t dt' \Gttprime^{(1)} (\xb) \rhoabs(t') \comma
                                                                \end{eq}
%-----------------------------------------------------------------------
where $\Gt^{(1)}(x) \equiv \int d\rt \Gt^{(1)}(\r,0)$ is the weighting
for a particle, initially at the interface, to be distant $x$ from the
interface after time $t$.  For arbitrary $z$, one just uses the
appropriate propagator $\Gt$ in eq. \eqref{integral-na-nb}.   

Before proceeding, let us use the above dynamics to prove eq.
\eqref{rate}, $\Rtdot=\lambda \rhoabs(t)$, a result that we have so
far assumed as physically obvious.  Now the total number of reactions
per unit area is $\Rt = \int d\xa [\nainf -\na(\xa;t)]$; integrating
the first of eqs. \eqref{integral-na-nb} over all $\xa$ and using the
fact that $\Gt^{(1)}(x)$ is normalized to unity, one has $\Rt =
\lambda \int_0^t dt' \rhoabs(t')$ which proves the desired result.

In the below we need calculate only one of the density fields, say the
less dense field $\na$, since one field uniquely implies the other.
This follows after subtracting the two equations in
\eqref{integral-na-nb}, and using $\Gt^{(1)}(x) = \Gt^{(1)}(-x)$, giving
%_______________________________________________________________________
                                                \begin{eq}{constraint}
\na(x;t) -\nb(-x;t) = \nainf - \nbinf \period
                                                                \end{eq}
%-----------------------------------------------------------------------
That is, the difference between mean A and B reactant densities at equal
distances from the interface is constant in time.

\subsection{Long Time Density at Interface}

It may appear that eq. \eqref{integral-na-nb} together with eqs.
\eqref{rate} and \eqref{pierre} provide a closed solution for the density fields.
However, in fact eq. \eqref{pierre} involves the unknown function
$\gamma(E)$, whose small $E$ behavior is needed to obtain the long
time density fields.  Hitherto we have asserted that its asymptotic
behavior is $\gamma(0)=1$, equivalent to the assertion that
$n(\infty)=\nbinf$ (see eqs. \eqref{n}, \eqref{pierre} and
\eqref{gamma} and surrounding discussions).  We must now prove these
assertions.  To do so, we will first argue that the A density at the
interface, $\nas(t) \equiv \na(0;t)$, vanishes for long times.  This
extra piece of information will allow the determination of
$\gamma(0)$.

We are able to prove $\nas(\infty)= 0$ by first relating $\nas$ to the
{\em like} particle correlation functions, $\rhoaas(t) \equiv
\rhoaa(0,0;t)$ and $\rhobbs(t) \equiv \rhobb(0,0;t)$, on the strength of
the following physically motivated assumption on these functions: 

\vi

{\bf Assumption 4.}
%_______________________________________________________________________
                                                \begin{eq}{ns-ass-one}
[\, \nas(t)\, ]^2 \leq  \rhoaas(t) \comma
                \gap [\, \nbs(t)\, ]^2 \leq  \rhobbs(t) \period
                                                                \end{eq}
%-----------------------------------------------------------------------

\vi

This states that reaction-induced correlations can only increase
density-density correlations of like particles, relative to the
totally random case where one would have $\rhoaas(t) = [\nas(t)]^2$.
That is, we admit the possibility of clustering of like particles.

To obtain information about $\rhoaas$ and $\rhobbs$ we first relate
them to $\rhoabs$.  In Appendix D we use Doi's framework to derive
dynamics for $\rhoaa,\rhobb$ from which we derive the following exact
equation
%_______________________________________________________________________
                                                \begin{eq}{monday}
\rhoaas(t) + \rhobbs(t) = (\nainf - \nbinf)^2 + 2\rhoabs(t) + 2\lambda \int_0^t
dt'  \Sd(t-t') \rhoabs(t')  \period
                                                                \end{eq}
%-----------------------------------------------------------------------
According to the results of sections 4,5 and 6, at long times
$\rhoabs(t)=\Rtdot/\lambda \twid t^{(1/z)-1}$.  (Note this conclusion
followed from assumption 3 and is quite independent of the numerical
value of $\gamma(0)$.)  Substituting this power law in eq.
\eqref{monday}, using $\Sd(t)\twid t^{-(d+1)/z}$ from eq.
\eqref{sd-def} and incorporating cut-offs in the marginal and
non-compact cases (see eqs. \eqref{cook}, \eqref{salt-on-food}
and \eqref{pepper-on-toast}) one sees that the time-dependent terms on
the right hand side of eq. \eqref{monday} tend to zero at long times.
Thus, making use of eq. \eqref{constraint}, we obtain from eq.
\eqref{monday}
%_______________________________________________________________________
                                                \begin{eq}{soliton}
[\, A(\infty) - 1\, ]\, [\, \nas(\infty)\, ]^2 + [\, B(\infty) - 1\, ]\, [\,
                                                        \nbs(\infty)\,]^2
                 = -2 \nas(\infty) \nbs(\infty) \comma 
                                                                \end{eq}
%-----------------------------------------------------------------------
where we have defined the unknown functions $A$ and $B$ such that
$\rhoaas(t) \equiv A(t) [\nas(t)]^2$ and $\rhobbs(t) \equiv B(t)
[\nbs(t)]^2$.  Note that assumptions 4 imply $A(t)$, $B(t) \geq 1$;
this in turn implies that the lhs of eq. \eqref{soliton} is positive
or zero.  But the rhs is negative or zero.  It follows that {\em both}
sides of this equation must vanish, \ie either or both of
$\nas(\infty)$ and $\nbs(\infty)$ vanish.  But from eq.
\eqref{constraint}, $\nas(\infty)\le \nbs(\infty)$.  Hence
$\nas(\infty) = 0$ is proved.

\subsection{Full Density Field}

Having determined that $\nas(\infty)$ vanishes, we return to eqs.
\eqref{integral-na-nb} from which we will first
determine $\gamma(0)$ and then calculate the full density profile.
Using the expression for $\Rtdot(E)$ in eq. \eqref{pierre}, and making
the substitution $\rhoabs(E) = \Rtdot(E)/\lambda$, eq.
\eqref{integral-na-nb} can be written in Laplace space as
%_______________________________________________________________________
                                                \begin{eq}{muffin}
\na(\xa;E) = {\nainf \over E} \square{ 1  - {\lambda \nbinf G^{(1)}_E (\xa) \over 1 +
\lambda \Sd(E) + \lambda \nbinf \gamma (E) \Sone(E)} } \period        
                                                                \end{eq}
%-----------------------------------------------------------------------
Here the Laplace transform of the propagator $\Gt^{(1)}(x)$ has the
following structure:
%_______________________________________________________________________
                                                \begin{eq}{donut}
G^{(1)}_E(x) = \Sone(E) \, {\widetilde g}(x E^{1/z}) \comma \gap
       {\widetilde g}(u) \gt \casesbracketsii{1}{u \ll 1}{0}{u \gg 1} \comma
                                                                \end{eq}
%-----------------------------------------------------------------------
where $\widetilde g$ is a scaling function with the stated limits.  We
have used eq. \eqref{green} and the fact (see eq. \eqref{sone}) that
$\Sone(t) = \Gt^{(1)}(x=0)$.

We can now prove $\gamma(E=0)=1$.  Consider the limit $E\gt 0$ of the
expression in eq. \eqref{muffin} evaluated at $\xa=0$.  In this limit
the square bracket must vanish since $\nas(t=\infty)=0$.  Now for
small enough $E$, the many body term $\lambda \nbinf \gamma (E)
\Sone(E)$ always dominates over the other two terms $1$ and
$\lambda\Sd(E)$ (see eqs. \eqref{food}, \eqref{terrible-confusion} and
\eqref{drink}).  Thus, using $G^{(1)}_E(0) = \Sone(E)$, 
we must have $\gamma(0)=1$.

Consider now general values of $\xa,t$ and let us compare the two
terms in the brackets on the rhs of eq. \eqref{muffin}.  According to
eq. \eqref{donut}, the numerator of the 2nd term is less than or equal
to the many body term in the denominator; it follows that the quotient
can be comparable to $1$ only for $E$ values sufficiently small that
the many body term dominates.  As we saw in eqs. \eqref{food},
\eqref{terrible-confusion} and \eqref{drink}, this corresponds to
times longer than the timescale signifying the crossover from second
to first order kinetics.  Therefore, retaining leading order terms
only in eq. \eqref{muffin}, one has
%_______________________________________________________________________
                                                \begin{eq}{hole}
\na(\xa;t) \approx \casesbracketsii
                {\nainf}{\mbox{``short'' times}}
                {\nainf \  f \paren{{\xa/ \xt}}\ \ } 
                {t \rightarrow \infty} 
                                                            \end{eq}
%-----------------------------------------------------------------------
where
%_______________________________________________________________________
                                                \begin{eq}{iced-coffee}
f \paren{\xa \over \xt} \equiv {\cal L}^{-1} \square{
                        1 -{\widetilde g}(\xa E^{1/z}) \over E} 
                                \comma \gap
f(u) \gt \casesbracketsii{0}{u \ll 1}
                          {1}{u \gg 1} 
                                                                \end{eq}
%-----------------------------------------------------------------------
and ${\cal L}^{-1}$ denotes inverse Laplace transform.  Here by
``short'' times, we refer to times when second-order kinetics are
valid.  This completes our calculation of the density profile, which
evidently confirms the physical expectations.  One sees that at short
times $\na(\xa;t)$ retains its equilibrium value, whereas at longer
times a reactant density depletion hole of size $\xt$ develops at the
interface (see fig. \ref{profile}).

As an example of the form of $f(u)$, consider small molecules ($z=2$)
for which $\Gt^{(1)}(x)$ is a Gaussian.  Determining $G_E^{(1)}(\xa)$,
one finds from eq. \eqref{iced-coffee} that $f(u) = {\rm Erf(u)}$;
this is identical to the asymptotic density profile in the situation
in which initially uniformly distributed small molecules adsorb
irreversibly onto a surface
\citeben{carslawjaeger:book,ben:adsorption}.

%*************************************************************************************
%*************************************************************************************
%*************************************************************************************
%*************************************************************************************

\section{Segregation Effects and Decay of Interfacial Density, $\nas(t)$.}

We found in Section 7 that the reactant density at the interface,
$\nas(t)$, tends to zero at long times.  In this section we determine
the long time power law decay of $\nas(t)$, considering for simplicity
the symmetric case only, $\nainf=\nbinf$.  Interestingly, we will find
segregation of reactants into A-rich and B-rich domains at the
interface for the compact case.

We begin by establishing time-dependent bounds on $\nas(t)$.  Now
assumption 4 will lead to an upper bound of this type, because for the
symmetric case $\rhoaas(t)$ can be determined from eq. \eqref{monday}
since $\rhoabs(t)$ is already known.  What we need, in addition, is a
lower time-dependent bound, which we now introduce by making one
further assumption.  This assumption is motivated by the physical
expectation that the density of A-B interfacial pairs, $\rhoabs(t)$,
will never exceed the value it would have if there were no
correlations between A and B particles, namely $[\nas(t)]^2$.  That
is, A-B reactions will always tend to diminish this pair density
relative to the uncorrelated value.

\vi

{\bf Assumption 5.} There exists a positive finite constant b, such that: 
%_______________________________________________________________________
                                                \begin{eq}{ns-ass-two}
b\, \rhoabs(t) \leq [\, \nas(t)\, ]^2  \gap (\nainf=\nbinf) \period
                                                                \end{eq}
%-----------------------------------------------------------------------

\vi

{\bf (a) Noncompact Case ($d+1>z$)}.  In Appendix D, eq.
\eqref{friday}, using eq. \eqref{monday} we show that for large
enough times $\rhoaas(t) \approx (\nainf/\lambda) d\xt/dt$.
Meanwhile, eqs. \eqref{noncompact-results} and
\eqref{noncompact-results-2} imply that $\rhoabs(t)
\approx (\nainf/\lambda) d\xt/dt$.  Thus the upper and lower long time
bounds on $\nas$ implied, respectively, by assumptions 4 and 5 are
proportional to the same algebraically decaying function of time.
Hence
%_______________________________________________________________________
                                                \begin{eq}{ns-non-compact}
\nas(t) \approx \sqrt{{\nainf \over \lambda}{d\xt \over dt}} \twid  t^{(1-z)/(2z)} 
\gap (t  \rightarrow \infty \comma \ d+1>z) \comma
                                                                \end{eq}
%-----------------------------------------------------------------------
up to a (time-dependent) prefactor of order unity. \\

{\bf (b) Compact Case ($d+1<z$)}.  For this case, as shown in appendix
D, eq. \eqref{monday} leads to the conclusion (see eq.
\eqref{saturday}) that asymptotically $\rhoaas(t) \approx \nainf
\xt^{-d} \twid t^{-d/z}$.  Meanwhile, eqs. \eqref{first-order-general} and
\eqref{skim-milk} imply the same decay for $\rhoabs$ as for the
noncompact case, $\rhoabs(t) \approx (\nainf/\lambda) d\xt/dt \twid
t^{(1-z)/z}$.  Hence the upper and lower bounds on $\nas(t)$ implied
by assumptions 4 and 5 involve different power laws: $\nas(t)$ decays
at least as slowly as $t^{-d/(2z)}$ and at least as rapidly as
$t^{(1-z)/(2z)}$.  This is insufficient to determine the actual decay.

We can make progress, however, by invoking the interface analogue of
the arguments which were used by Ovchinnikov and Zeldovich
\citeben{ovchinnikovzeldovich:segregation}, and Toussaint and Wilczek
\citeben{toussaintwilczek:segregation} to analyze the bulk reaction
system $A+B\gt 0$.  According to this generalization, which we have
presented in the introduction, the density of A reactants at the
interface cannot decay faster than $\sqrt{\nainf} \xt^{-d/2}
\twid t^{-d/(2z)}$, which is the rate determined by the decay of
fluctuations in the random initial reactant distribution.  But we have
already shown that  $ t^{-d/(2z)}$ is an upper bound.  Hence
%_______________________________________________________________________
                                                \begin{eq}{ns-compact}
\nas(t) \approx \sqrt{\nainf} \xt^{-d/2} \twid t^{-d/(2z)}  \gap
                                  (t \rightarrow \infty \comma d+1<z)\period
                                                                \end{eq}
%-----------------------------------------------------------------------
Therefore, the asymptotic density decay at the interface is controlled
by the rate of decay of fluctuations.  It follows that $A$-rich and
$B$-rich regions of linear size $\xt$ develop adjacent to the
interface.  These are illustrated schematically in fig.
\ref{segregation}.  An important point to stress is that the long time reaction
rate is itself {\em not} influenced by this segregation, to leading
order: the long time reaction rate is governed merely by the fact that
$\nas(\infty)=0$, whilst segregation effects are associated with
higher order terms in $\nas(t)$, \ie the manner in which $\nas$ decays
to zero.

%************************************************************************************
%************************************************************************************
%************************************************************************************
%************************************************************************************

\section{Discussion}

We have shown here that the critical dimension for reaction kinetics
at a fixed interface is $d_c=z-1$.  This is quite different to the
result for reactions at a movable and broadening interface separating
2 {\em miscible} phases, which problem has been widely studied
for the case $z=2$ where $d_c=2$ has been
found \citeben{cornelldrozchopard:aplusbfront_fluc,%
cornelldroz:aplusb_front_prl,leecardy:aplusb_front,%
howardcardy:aplusb_front_rg}.  For the fixed interface problem studied
here, one has instead $d_c=1$.  The difference between these 2
critical dimensions is due to the fact that for the case of miscible
reactants, A-B reactions are not restricted to occur only in a
$(d-1)$-dimensional plane.

The most novel feature to have emerged from this study is that
interfacial reaction kinetics are not of fixed order.  This is rather
unusual.  For example, trimolecular, bimolecular and unimolecular
reaction processes are generally governed by 3rd, 2nd and 1st order
kinetics, respectively.  The peculiar feature here is that 2nd order
reaction rate laws are obeyed at short times, whilst 1st order
kinetics describe the long time behavior:  
%_______________________________________________________________________
                                                \begin{eq}{salt-on-lips}
\Rtdot = \ksecond \nainf \nbinf \ \ (\mbox{2nd order})\comma\gap
\Rtdot = \kfirst \nainf\ \  (\mbox{1st order})\period
                                                                \end{eq}
%-----------------------------------------------------------------------

The 2nd order coefficient $\ksecond$ may either be a constant (mean
field kinetics, MF) or time-dependent (diffusion-controlled kinetics,
DC).  The time-dependence in the latter case is $\ksecond \approx
d\xt^{d+1}/dt$.  In contrast, the 1st order kinetics are always DC,
and the 1st order coefficient $\kfirst \approx d\xt/dt$ is always
time-dependent.  

An important feature of these 1st order kinetics concerns the
different roles played by the two far-field reactant densities,
$\nainf$ and $\nbinf$, in the case where they are unequal.  The
timescale at which these kinetics onset (either $\tstarmany$ or $\tl$)
is determined by the {\em greatest}, $\nbinf$.  However, the rate law
itself involves the {\em smallest} one, $\nainf$ (see eq.
\eqref{salt-on-lips}).  Correspondingly, in the region within a
distance $\xt$ of the interface the density profile falls to a value
close to zero on the dilute A side, whereas on the denser B side the
profile in this region drops to a finite value close to
$\nbinf-\nainf$.

Apart from our main concern, the reaction rate, this paper has also
addressed the evolution of density fields, key features of which are
the densities at the interface $\nas,\nbs$.  This enabled us to
examine the validity of the ``local mean field'' decoupling
approximation, $\Rtdot \approx \lambda \nbs(t)\, \nas(t)$, which approximates
the densities on either side of the interface to be independent of one
another, $\rhoabs\approx \nas \nbs$.  In this picture the denser B
side is viewed as presenting a uniform ``reactive surface'' of strength
$\lambda
\nbs(t)$ to the more dilute A reactants.  Consider long times, when
the diffusive flux of A particles per unit area at the interface is
$\nainf\, d \xt /dt$.  Equating this to the reaction rate, one sees
that the ``local mean field'' approximation suggests
%_______________________________________________________________________
                                                   \begin{eq}{yesterday}
\nas(t) \approx \inverse{\lambda \nbs(t)} {d\xt \over dt} \nainf  
          \gap (t \gt \infty\comma \ \mbox{local mean field approx.}) 
                                                                \end{eq}
%-----------------------------------------------------------------------
Now in the symmetric case, $\nas=\nbs$, the above result implies
$\nas(t) \twid t^{(1-z)/(2z)}$.  But we saw in section 8 that the decay
rate is always limited by the rate at which fluctuations in the
initial differences between densities on the A and B side near the
interface can diffuse away.  This limiting decay was shown to be
$\twid t^{-d/(2z)}$.  This suggests that only in high dimensions,
$d+1>z$, is the $\nas(t) \twid t^{(1-z)/(2z)}$ prediction correct;
indeed, we demonstrated this in section 8.  We conclude that the local
mean field approximation is essentially valid for $d+1>z$ at very long
times.  For all low dimensions $d+1<z$, however, fluctuations
determine the decay law: $\nas(t) \twid t^{-d/(2z)}$, segregation
occurs at the interface, $\nas$ and $\nbs$ are no longer independent
and the the local mean-field approximation is wrong.

Let us make a few comments about the interfacial densities in the
asymmetric case, $\nbs > \nas$.  In this case we expect eq.
\eqref{yesterday} to be valid for {\em all} dimensions, since
$\nbs(\infty) = \nbinf - \nainf$ is then non-vanishing (see eq.
\eqref{constraint}).  Hence the B side will indeed supply a
uniform reactive surface for the A reactants.  Thus, we expect a
different decay law, $\nas \twid t^{(1-z)/z}$ for all dimensions.  
If the initial reactant densities $\nainf, \nbinf$ are
almost but not quite equal to one another, we expect the symmetric
case results will be valid up to a cross-over time at which $\nbs(t)$
drops to a value close to its asymptotic value, $\nbinf - \nainf$.
Thereafter, eq. \eqref{yesterday} will correctly describe $\nas$.

We stress that this study has concerned {\em irreversible} reactions.
Thus an equilibrium state is never attained.  The final state will be
governed by saturation effects at the interface, which have not been
considered here.  As $t \rightarrow \infty$, in principle a final
state will be attained in which reaction product fills every available
surface site.  (In practice, however, the timescale for this state to
be reached may be experimentally inaccessible
\citeben{ben:reactiface_pol_letter,ben:reactiface_pol}).

To conclude, consider a few specific examples.  An important parameter
determining the class of reaction kinetics is the dimensionless local
reactivity, $\Qbare \ta$.  Perhaps the most useful relation to help one estimate
its value for a given system is $\Qbare \ta \approx
\kbulk/\kbulkradical$ where $\kbulk\approx \Qbare a^3$ is the 
bulk rate constant, \ie the 3-dimensional rate constant which would
describe A-B reaction kinetics if the molecules could react anywhere
within the bulk (see introduction).  Here $\kbulkradical \approx
a^3/\ta \approx 10^9$(litres/mol)sec$^{-1}$ is the same quantity
for radicals which are nature's most reactive chemical species.  We
assume here the molecular size $a$ is roughly the same ($a\approx
3$\Angstrom) in all small molecule cases.  Thus if one has access to
$\kbulk$ for an A-B system, then one can estimate $\Qbare \ta$.  In
the case where the reactive groups are attached to polymer chains,
$\kbulk$ refers of course to the {\em small molecule} bulk analogue
reaction system, \ie the rate constant describing reactions between
the same species after removal from their host polymer chains.

\subsection*{Small molecules: $z=2$, $d=3$.}

Consider firstly unequal initial bulk densities, $\nainf\ne\nbinf$.
The early 2nd order behavior is non-compact ($d+1>z$) and MF 2nd order
kinetics pertain with $\ksecond=h (\Qbare a^3) $.  These continue
until $\tstarmany = D/[h (\Qbare a^3) \nbinf ]^2$, when first order
kinetics onset with time-dependent rate constant $\kfirst =
D/(Dt)^{1/2}$ where $D$ is the molecular diffusivity.  Note that the
cross-over time $\tstarmany$ is determined by the greater of the two
far-field densities, $\nbinf$.

The density profile on the less dense A side is (to leading order)
identical to the ``reactive surface'' situation, having a depletion
hole of size $(Dt)^{1/2}$.  The A density at the interface decays for
long times to zero as $\nas\twid 1/t^{1/2}$.  There is no ``hole'' on
the more dense B side, though the density is reduced from its initial
value over a region extending $(Dt)^{1/2}$ into the bulk and has the
long time value $\nbinf-\nainf$ at the interface.  The symmetric case,
$\nainf=\nbinf$, is different: there are long time holes on both sides
and the interfacial density decay is $\nas\twid 1/t^{1/4}$.

Typical numerical values are $a\approx h \approx 3$\Angstrom\ and
$D\approx 10^{-5}$cm$^2/$sec.  Now for the vast majority of reacting
species, $\kbulk\lsim 10^{3}$ (litres/mol)sec$^{-1}$, implying
$\tstarmany\gsim 10\ {\rm sec}/\phi_B^2$, where $\phiB = \nbinf a^3$
is the far-field volume fraction of B reactants.  Thus, depending on
the value of $\phi_B$, this timescale may become so large that the
diffusion-controlled kinetics get washed out by other effects such as
convection.  For highly reactive species such as radicals, on the
other hand, one has $\kbulk\approx10^9$(litres/mol)sec$^{-1}$ and
$\tstarmany \approx 10^{-10} {\rm sec} /\phi_B^2$; these kinetics are
then observable over a very large range of densities.

\subsection*{Small molecules in $d=1$.} 
This is a marginal situation ($z=d+1$) arising in systems where small
molecules ($z=2$) are restricted to an effectively one-dimensional
geometry, \eg molecules trapped in a thin tube.

For highly reactive species, $\Qbare\ta\approx 1$ (\ie $\kbulk\approx
\kbulkradical$) the initial regime is
2nd order with a weakly time-dependent rate constant $\ksecond \approx
D/\ln(t/\th)$.  At time $\Tl = D^{-1} (\nbinf)^{-2} [\, \ln (\nbinf
h)^2\,]^2$ first order kinetics onset with time-dependent $\kfirst =
D/(Dt)^{1/2}$.

For most cases, however, the local dimensionless reactivity
$\Qbare\ta$ will be below a very high threshold value (\ie very close
to unity) given by $\Qbare^*\ta= (a/h)/[\,\ln(\nbinf h)^2\,]^2$.  In such cases
an initial 2nd order mean field regime with $\ksecond = \Qbare h a =
\kbulk h/a^2$ is followed at time $\tstarmany = \ta/(\Qbare\ta
\nbinf h)^2$ by the same 1st order kinetics.  In all cases a depletion
hole of size $(D t)^{1/2}$ grows at long times on the dilute A side.

\subsection*{Unentangled polymers, short times: $z=4, d=3$.} 
Consider an interface separating two immiscible unentangled polymer
melts comprising chains with degree of polymerization $N$ and radius
of gyration $R$, each carrying one reactive group.  Thus the density
of reactive groups is $\nainf=\nbinf = 1/(N a^3)$ or equivalently
$\phi_A = \phiB = 1/N$.  Then Rouse dynamics apply
\citeben{doiedwards:book,gennes:book}, $\xt \approx R (t/\tau)^{1/4}$ for
times less than the single chain longest relaxation time, $\tau
\approx \ta N^2$.  Thus, for $t < \tau$, we are in the marginal
situation $z=d+1=4$.

Consider first the maximally reactive case $\Qbare \ta = 1$. Initially
kinetics are 2nd order with weakly time-dependent rate constant
$\ksecond = (R^4/\tau)/\ln(t/\th)$.  The cross over to first order
kinetics, with $\kfirst \approx (R/\tau)\, (t/\tau)^{-3/4}$, occurs at
$\Tl = \ta \phiB^{-4/3} \ln(\phiB^{-4/3}a^4/h^4)$.  For typical values
$h/a=5$, $N = 200$ one has $\Tl \approx 0.02 \, \tau$.

For less reactive species, $\Qbare \ta <\Qbare^*\ta$ where $\Qbare^*
\ta \approx a/[h\ln(\phiB^{-4/3} a^4/h^4)]$, the kinetics are
different.  (For the above typical numerical values, $\Qbare^*
\ta \approx 0.7$.)  In this case second order MF kinetics with $\ksecond=h (\Qbare
a^3)$ are followed at $\tstarmany = \ta [\Qbare (h/a) \ta
\phiB]^{-4/3}$ by 1st order kinetics with $\kfirst = (R/\tau)
(t/\tau)^{-3/4}$.

In both of these examples, a long time depletion hole grows on the
dilute A side whose size increases in time as $\xt \approx
R(t/\tau)^{1/4}$.

\subsection*{Entangled polymers, ``breathing'' modes: $z=8, d=3$.}
Consider the same polymer example as above, but now chains are
entangled.  Using the reptation model to describe the polymer
dynamics, let us ask what reaction kinetics are during the short time
``breathing modes'' regime ($\te<t<\tb$) when
\citeben{doiedwards:book,gennes:book} $\xt = r_{\rm e}(t/\te)^{1/8}$.
Here $\te = \Ne^2 \ta$ is the entanglement time ($\Ne$ being the
entanglement threshold), $\tb = (N/\Ne)^2 \te$ is the Rouse time for
the one-dimensional tube motion and $r_{\rm e}=\Ne^{1/2}a$ is the
tube diameter.  This is an interesting example of a compact case,
$d+1<z=8$.

Consider very reactive groups such as radicals, $\Qbare\ta\approx 1$,
and $\nbinf$ values such that $\tl$ (the diffusion time corresponding
to a distance equal to the typical separation between the B reactive
groups) satisfies $\te < \tl <\tb$.  Then $\tl = \te (\nbinf r_{\rm
e}^3)^{-8/3}$.  Now for $t>\te$ there is no MF regime and kinetics
are 2nd order DC, $\ksecond \approx (r_{\rm e}^4/\te) (t/\te)^{-1/2}$.
For $t>\tl$ the kinetics become 1st order with $\kfirst = (r_{\rm
e}/\te) (t/\te)^{-7/8}$ and a depletion hole grows on the A side of
size $\twid t^{1/8}$.

As a specific example, if all B-chains carry one
reactive end-group ($\nbinf a^3 = 1/N$) and $N \approx 10^4$, $\Ne
\approx 200$ one has $\tl \approx 30 \te \approx 10^{-3} \tb$. Thus 
both 2nd and 1st order kinetics as described above will occur within
the $t^{1/8}$ regime.

%************************************************************************************
\vii

{\small

This work was supported by the National Science Foundation under grant
no. DMR-9403566.  We thank Uday Sawhney for stimulating discussions.

}

%************************************************************************************
%************************************************************************************
%**************************** START APPENDIX ****************************************
%************************************************************************************

{\appendix

\section{Derivation for $z=2$ of Dynamical Equations for 2-Body and 1-Body
Density Correlation Functions.}

In this Appendix we employ the second-quantization formalism for
classical many-particle systems developed by Doi
\citeben{doi:reaction_secondquant1and2} to derive exact evolution
equations, in the case of small molecules ($z=2$), obeyed by the
2-body correlation functions $\rhoab, \rhoaa, \rhobb$ and the density
fields $\na, \nb$.  We do not attempt here a self-contained
discussion of the Doi formalism: the reader is referred to Doi's
papers, refs.  \citenum{doi:reaction_secondquant1and2}, for the
necessary background.

In the Doi formalism any physical quantity $A$ is mapped onto a
quantum operator $\widetilde{A}$ given in terms of the Bose creation
and annihilation operators.  For systems consisting of two types of
particles, A and B, there are two types of Bose operators, $\psia(\r),
\psib(\r)$, for every spatial location $\r$.  These satisfy the
following commutation relations
%_______________________________________________________________________
                                                \begin{eq}{bose}
[\psinu(\r),\psimudag(\rprime)] = \delta(\r-\rprime) \delta_{\nu \mu}
\comma \gap 
[\psinudag(\r),\psimudag(\rprime)] = [\psinu(\r),\psimu(\rprime)]=0 
\comma
                                                                \end{eq}
%-----------------------------------------------------------------------
where $\nu=A,B$ and $\mu=A,B$.  The dynamics of $A(t)$ are determined
by a quantum propagator $\Gtwiddle$,
%_______________________________________________________________________
                                                \begin{eq}{inner}
A(t) = \left< 1 \left| {\widetilde A} e^{- \Gtwiddle t} \right|c \right> 
                                         \period
                                                                \end{eq}
%-----------------------------------------------------------------------
Here, $\left< 1 \right| \equiv \left<0 \right| \exp [\int d\ra
\psia(\ra) \int d\rb \psib(\rb)]$ is a coherent state,
 $\left< 0 \right| $ is the vacuum state and $\left<1 \right|
\psiadag(\r) = \left<1 \right| \psibdag(\r) = \left< 1
\right|\,$.  The quantum state $\left| c \right>$ represents the
initial state of the system; although its form will not be relevant in
the subsequent calculations, we remark that in the case where A and B
particles are initially randomly distributed with densities $\nainf$
and $\nbinf$ then $\left| c \right> = \exp[\nainf \nbinf
\int d\ra \psiadag(\ra) \int d\rb \psibdag(\rb)] \left| 0 \right>$.  

In the present reaction-diffusion interface problem, the propagator
$\Gtwiddle = \Gtwiddle_0 + \Gtwiddle_{\rm r}$ consists of a
``diffusion'' part, $\Gtwiddle_0$, and a ``reaction'' part,
$\Gtwiddle_{\rm r}$.  For small molecules obeying simple Fickian
diffusion, according to Doi
\citeben{doi:reaction_secondquant1and2}
%_______________________________________________________________________
                                                \begin{eq}{gzero}
\Gtwiddle_0 = - D \int d\ra \psiadag(\ra) \nabla^2_A \psia(\ra) -
                D \int d\rb \psibdag(\rb) \nabla^2_B \psib(\rb)  \period
                                                 \end{eq}
%-----------------------------------------------------------------------
The reaction part, $\Gtwiddle_{\rm r}$, is constructed
\citeben{doi:reaction_secondquant1and2} from the reaction sink
function which in our model is $\lambda \delta(\xa) \delta(\ra -
\rb)$.  We find that the corresponding quantum operator is
%_______________________________________________________________________
                                                \begin{eq}{greaction}
\Gtwiddle_{\rm r} = \lambda \int d\rt
\curly{\psiadag(\rt)\psibdag(\rt)\psia(\rt)\psib(\rt) -
\psia(\rt)\psib(\rt)} \period
                                                                \end{eq}
%-----------------------------------------------------------------------
The two annihilation operators in the second term in $\Gtwiddle_{\rm
r}$  are responsible for reactions between A-B pairs in contact at the
interface, while the first part is necessary to ensure proper
normalization of averages\citeben{doi:reaction_secondquant1and2}.

Differentiating eq. \eqref{inner}, and using the identity $\left< 1
\right| \Gtwiddle = 0$ which follows from the above definition of
$\Gtwiddle$, one has
%_______________________________________________________________________
                                                \begin{eq}{quantum-evolution}
{d A(t) \over dt} = \left< 1 \left| [\Gtwiddle, {\widetilde A}]
                                e^{-\Gtwiddle t} \right| c \right> \period
                                                                \end{eq}
%-----------------------------------------------------------------------
We can now derive the dynamical equations obeyed by the correlation
functions from eq. \eqref{quantum-evolution}.  The operator
representations of the many-body correlation functions are
\citeben{doi:reaction_secondquant1and2} as follows:
%_______________________________________________________________________
            \begin{eqarray}{doi-rep}
\ntwiddle_{\nu}(\r)  & =& \psinudag(\r) \psinu(\r) \comma
                                               \drop
{\widetilde \rho}_{\mu \nu} (\r,\rprime) & = & \psinudag(\r)
\psimudag(\rprime) \psinu(\r) \psimu(\rprime)  \comma
                                                 \drop
{\widetilde \rho}_{\sigma \mu \nu}(\r, \rprime, \rprime') &=&
\psi_{\sigma}^{\dag}(\r) \psimudag(\rprime) \psinudag(\rprime')
\psi_{\sigma}(\r) \psimu(\rprime) \psinu(\rprime')  \gap
(\sigma,\mu,\nu = A,B) \period  \drop
                                                                \end{eqarray}
%-----------------------------------------------------------------------
From eq. \eqref{quantum-evolution}, with $A=\na$ and ${\widetilde
A}=\ntwiddle_A$, and using the above representation for $\ntwiddle_A$,
one obtains
%_______________________________________________________________________
                                                \begin{eq}{dream}
{d \na(\ra;t)\over dt }= D \nabla_A^2 \left< 1 \right|
\psiadag (\ra) \psia(\ra) e^{-\Gtwiddle t} \left| c \right> - \lambda \delta(\xa)
\left< 1 \right| \psiadag(\rat)
\psibdag(\rat) \psia(\rat) \psib(\rat) e^{-\Gtwiddle t} \left| c \right>
                                                                \end{eq}
%-----------------------------------------------------------------------
after using the commutation relations of eq. \eqref{bose} and the
properties of the coherent state $\left<1\right|$.  From eqs.
\eqref{doi-rep} and eq. \eqref{inner} one recognizes this as  the
first of eqs. \eqref{na-nb} in the main text (dynamics of $\na, \nb$).
Notice that the commutation of ${\Gtwiddle}$ with $\ntwiddle$ in eq.
\eqref{quantum-evolution} produced, among other quantities, a higher
order correlation function; this is the origin of the hierarchical
structure of the reaction-diffusion equations.

Performing a similar analysis for the 2-body correlation functions
(setting $A=\rho_{\mu \nu}$ and ${\widetilde A}={\widetilde
\rho}_{\mu \nu}$) one obtains eqs. \eqref{rhoab-dot} (dynamics of
$\rhoab$) and eq. \eqref{honey} (dynamics of $\rhoaa, \rhobb$).

%************************************************************************************
%************************************************************************************

\section{Proof That Relative Contribution of Region I to $\Ima$ is
Order Unity} 

We saw in section 3 that the many-body integral term $\Ima$ involving
$\rhobab(\r|0,0;t)$ in eq. \eqref{broom} receives contributions from
two space-time regions, I and II.  In this appendix we demonstrate
that the contribution from region I is a fraction of order unity of
the value of $\Ima$ itself.  We show this by firstly deriving an upper
bound on $\Ima$.  Then we derive a lower bound on $\Ima$ corresponding
to $\rhobab$ being zero in region II and having its minimum value in
region I.  The lower and upper bounds will then be shown to be of the
same order, proving the desired result.  These bounds are all
consequences of assumptions 1, 2 and 3 (see main text).

From assumption 1 one sees that the substitution $\rhobab(\rb'|0,0;t)
\gt U \nbinf$ in the expression for $\Ima$ (eq. \eqref{broom}) defines an upper
bound on $\Ima$,
%_______________________________________________________________________
                                                \begin{eq}{supper}
\Ima \leq U \nbinf \int_0^t dt' \Sone(t-t') \rhoabs(t') \comma
                                                                \end{eq}
%-----------------------------------------------------------------------
where we have used eq. \eqref{green-product} to perform the integration.

Now assumption 2 implies that
%_______________________________________________________________________
             \begin{eq}{rho-min}
\rhobab(\rbprime | 0,0;t') \geq
\casesbracketsii{L \nbinf}{\xb' / x_{t'}>1\comma\  \mbox{region I}}
                      {0}{\xb' / x_{t'} <1 \comma\  \mbox{region II}}  \period
            \end{eq}
%-----------------------------------------------------------------------
Substituting the above lower bound on $\rhobab$ in the expression
for $\Ima$ of eq. \eqref{broom}, we obtain the following lower bound
on $\Ima$
%_______________________________________________________________________
                                                \begin{eq}{ima-lower}
\Ima \geq
  L \nbinf \int_0^t dt' \,{1 \over x_{t-t'}} \ \rhoabs(t') 
                \int_{\xi_x > (x_{t'}/x_{t-t'})}
                 d^d \xi \  h (\xi)  \comma 
                                                                \end{eq}
%-----------------------------------------------------------------------
where we have used the following scaling structure for the function
$F_t(\rb')$ appearing in the expression for $\Ima$ of eq. \eqref{broom},
%_______________________________________________________________________
                                                \begin{eq}{f-scaling}
F_t (\rb' ) = {1 \over \xt^{d+1}} \, h \paren{\rb' \over \xt} 
        \comma \gap 
h(\bu)  \gt  \casesbracketsshortii{1}{u \ll1}
                                      {0}{u \gg 1}
                                                                \end{eq}
%-----------------------------------------------------------------------
This follows from eq. \eqref{green}.  The integration variable in eq.
\eqref{ima-lower} is $\xi=\rb'/x_{t-t'}$ and $\xi_x$ denotes the
component of $\xi$ orthogonal to the interface.

Now using eq. \eqref{f-scaling}, there exists a positive constant
$E$ of order unity such that
%_______________________________________________________________________
                                                \begin{eq}{f-inequality}
\int_{\xi_x > (x_{t'} / x_{t-t'})} d^d \xi \, h (\xi)\,  \geq
        \casesbracketsii{E}{{x_{t'} / x_{t-t'}} < 1 \ {\rm or} \  t' < {t / 2}}
                     {0}{{x_{t'} / x_{t-t'}} > 1 \ {\rm or} \  t' > {t / 2}}
                                                                \end{eq}
%-----------------------------------------------------------------------
Expression \eqref{f-inequality} in inequality \eqref{ima-lower} implies that 
%_______________________________________________________________________
                                                \begin{eq}{ima-lower-2}
\Ima \geq E L \nbinf \int_0^{t/2} dt' {1 \over x_{t-t'}} \rhoabs(t') \period
                                                                \end{eq}
%-----------------------------------------------------------------------
Notice that according to eq. \eqref{sd-def}, $\Sone(t) \approx 1/\xt$.
Therefore, inequality \eqref{ima-lower-2} is very close to showing
that the lower bound on $\Ima$ is of the same order as the upper bound
in eq. \eqref{supper}, except for the fact that the time integral on
the right hand side of this inequality has upper limit $t / 2$ rather
than $t$.  However, if one makes the replacement $t/2\gt t$ for this
upper limit, this yields the same result to within a constant of order
unity.  This is a consequence firstly of the fact that since $z>1$ in
eq. \eqref{xt} thus both $\int_0^{t/2} dt'/x_{t-t'}$ and
$\int_{t/2}^{t} dt'/x_{t-t'}$ are of the same order, and secondly that
$\rhoabs(t')$, according to assumption 3, is a decreasing function of
time.

Thus we have shown that the upper and lower bounds on $\Ima$ are of
the same order.  But the crucial point is that the lower bound on $\Ima$ of eq.
\eqref{ima-lower} results from an integration receiving {\em zero}
contribution from region II and minimal contribution from region I.
It follows that the actual contribution from region I must be of the
same order as the actual value of $\Ima$.

%************************************************************************************
%************************************************************************************

\section{Inverse Laplace Transform of $1/(E\ln E)$}

We show in this Appendix that 
%_______________________________________________________________________
                                                \begin{eq}{final-pair}
{\cal L}^{-1}\square{-1 \over E \ln (E \th)} = {1 \over \Gamma(1) \ln(t/\th)} 
                 \gap  (t \gg \th) \comma
                                                                \end{eq}
%-----------------------------------------------------------------------
where ${\cal L}^{-1}$ denotes inverse Laplace transform.
Now it is well known that 
%_______________________________________________________________________
                                                \begin{eq}{power-pair}
{\cal L}^{-1} \square{{1 \over (E\th)^n}} = {t^{n-1} \over \th^{n} \Gamma(n)}  \gap
(n>0) \period
                                                                \end{eq}
%-----------------------------------------------------------------------
Integrating both sides of eq. \eqref{power-pair} with respect to $n$
from $n=0$ to $n=1$ then gives
%_______________________________________________________________________
                                                \begin{eq}{integral-pair}
{\cal L}^{-1} \square{\paren{1-{1 \over E \th}} {1 \over \ln E \th}}
= \inverse{t} \ \int_0^1 { e^{ n \ln (t/\th)} \over \Gamma(n)} dn \period
                                                                \end{eq}
%-----------------------------------------------------------------------
The function $1/\Gamma(n)$ is finite for all $n$ in $[0,1]$. Thus,
expanding $1/\Gamma(n)$ in a Taylor series around $n=1$ in the
integral on the rhs of eq. \eqref{integral-pair}, we obtain
%_______________________________________________________________________
                                                \begin{eq}{expansion}
\inverse{t} \ \int_0^1 { e^{ n \ln (t/\th)} \over \Gamma(n)} dn = 
\inverse{\Gamma(1) \th \ln(t/\th)}  + O\paren{ \inverse{\th [\ln(t/\th)]^2} 
}  
 \gap ( t \gg \th) \period
                                                                \end{eq}
%----------------------------------------------------------------------
Considering the limit $t \gg \th$, corresponding to $E\th \ll 1$, from
eqs. \eqref{integral-pair} and \eqref{expansion} we deduce eq.
\eqref{final-pair}.

%************************************************************************************
%************************************************************************************

\section{Like Particle Correlation Functions $\rhoaa, \rhobb$: 
Asymptotic Decay and Proof of Eq. (55)} 

Using Doi's second-quantization formalism
\citeben{doi:reaction_secondquant1and2} it is shown in Appendix A
that for $z=2$ (small molecules) $\rhoaa(\ra,\ra';t)$ and
$\rhobb(\rb,\rb';t)$ obey the following dynamics
%_______________________________________________________________________
                                        \begin{eqarray}{honey}
\curly{
{\partial \over \partial t}  
- D\,[\nabla_{\ra}^2 + \nabla_{\raprime}^2]
}
\rhoaa(\ra,\raprime;t)    & = & 
                  - \lambda \delta (\xa) \rhoaab(\ra,\raprime,\ra;t) -
                    \lambda \delta(\xa') \rhoaab(\ra,\raprime,\raprime;t)
                                        \drop
\curly{
{\partial \over \partial t}  
- D\,[\nabla_{\rb}^2 + \nabla_{\rbprime}^2]
}
\rhobb(\rb,\rbprime;t)    & = &
                  - \lambda \delta (\xb) \rhobba(\rb,\rbprime,\rb;t) -
                    \lambda \delta(\xb') \rhobba(\rb,\rbprime,\rbprime;t) \period
                                                \drop
                         \end{eqarray} 
%-----------------------------------------------------------------------
Eqs. \eqref{honey} have a similar form to eq. \eqref{rhoab-dot} for
the $\rhoab(\ra,\rb;t)$ dynamics except that the two-body sink term
responsible for pair reactions in eq. \eqref{rhoab-dot} is absent
since two particles of the same species cannot react with one another.
Solving eqs. \eqref{honey} and setting $\ra,\ra',\rb,\rb'=0$ we obtain
%_______________________________________________________________________
                   \begin{eq}{rhoaa-rhobb-integral}
\rhoaas(t)    =  (\nainf)^{2} - 2 \lambda \Imb \comma \gap
\rhobbs(t)    =  (\nbinf)^{2} - 2 \lambda \Ima  \comma
                                           \end{eq}
%-----------------------------------------------------------------------
where $\Ima, \Imb$ are defined in eq. \eqref{im-def} of section 2 and
are the identical many-body integral expressions which appeared in the
solution for $\rhoabs$ of eq. \eqref{exact-rhoabs}.

Using the above expressions for the like interfacial pair densities
together with eq. \eqref{exact-rhoabs} for $\rhoabs$, one obtains the
exact relation displayed in eq. \eqref{monday}.  This result tells us
we can determine the sum $\rhoaas(t)+\rhobbs(t)$ from knowledge of
$\rhoabs(t)$.  For general $z$, eq. \eqref{monday} remains valid,
provided one replaces everywhere the Gaussian $z=2$ propagator with
the propagator $\Gt$ appropriate to the dynamics.

Consider the symmetric case, $\rhoaas(t)= \rhobbs(t)$.  Laplace
transforming eq. \eqref{monday} we obtain
%_______________________________________________________________________
                                                \begin{eq}{thursday}
\rhoaas(E) = \rhoabs(E) \curly{1 + \lambda \Sd(E)}   \gap (\nainf = \nbinf). 
                                                                \end{eq}
%-----------------------------------------------------------------------
In the non-compact case ($d+1>z$), from eqs. \eqref{salt-on-food}
and \eqref{pepper-on-toast}, according to which $\Sd(E)$ is a constant,
and from the asymptotic form of $\rhoabs(t) = \Rtdot/\lambda \twid
t^{1/z-1}$ (see eqs. \eqref{noncompact-results} and \eqref{noncompact-results-2}) one has
%_______________________________________________________________________
                                                \begin{eq}{friday}
\rhoaas(t) \approx \casesbracketsii
                           { \{(1+\lambda \th/h^{d+1}) / \lambda\} \
                                        \nainf\, d\xt/dt}
                                                {z<d+1<z+1}
                             { \{ (1 + \Qbare \ta) / \lambda \}\ \nainf\, d\xt/dt}
                                                {d>z}
\ \ \ (\nainf = \nbinf, \ t \gt \infty)
                                                   \end{eq}
%-----------------------------------------------------------------------
Meanwhile in the compact case, from eq. \eqref{sd-def} one has $\Sd(E)
\twid E^{(d+1)/z-1}$; thus, using $\rhoabs(E) = \Rtdot(E)/\lambda \twid
E^{-1/z}$ from eq. \eqref{food}, valid for $E \gt 0$, one has from eq.
\eqref{thursday}
%_______________________________________________________________________
                                                \begin{eq}{saturday}
\rhoaas(t) \approx \nainf\, \xt^{-d} \twid t^{-d/z} \gap 
                                        (d<z\comma\ \nainf = \nbinf, \ t \gt
\infty) \period 
                                                                \end{eq}
%-----------------------------------------------------------------------

%*************************************************************************************
\ignore{
From the results of
Sections 4, 5 and 6 for the long time reaction rate, we have that
$\rhoabs(t) \approx (\nainf/\lambda) t^{(1/z) -1 }$ asymptotically.

Using the fact that $\Sd(t) \approx t^{-(d+1)/z}$, we have from the
definition of $I(t)$ in eq. \eqref{monday} in the compact case,
%_______________________________________________________________________
                                                \begin{eq}{i2-compact}
I(t) \approx {\nainf \over \lambda} t^{-d/z} \int_0^1 d \xi {\xi
^{1/z -1} \over (1-\xi)^{(d+1)/z}} \approx {\nainf \over \lambda}
t^{-d/z} \comma \gap (z > d+ 1) \period 
                                                                \end{eq}
%-----------------------------------------------------------------------

In the non-compact case, the integral is dominated by $t' \approx t$.
Using the fact that $\Sd(t) \approx 1/h^{d+1}$, for $t \leq \th$, we
have:
%_______________________________________________________________________
                                                \begin{eq}{i2-non-compact}
I(t) \approx {\nainf \over \lambda} \int_{t-\th}^{t} dt'
{{\tprime^{(1/z)-1}} \over h^{d+1}} \approx {\nainf \bar{Q} \th \over
\lambda^2} t^{1/z -1} \comma \gap (z < d+1) \period
                                                                \end{eq}
%-----------------------------------------------------------------------

In the marginal case we have 
%\citeben{abramowitzstegun:math_tables}:
%_______________________________________________________________________
                                                \begin{eqarray}{i2-marginal}
I(t) \approx {\nainf \over \lambda} \int_0^{t- \th} dt' {\ {\tprime}^{(1/z)-1}
\over t-t'} &\approx& {\nainf \over \lambda} t^{(1/z)-1} F(1,1/z,1+1/z,1-\th/t) 
                                                        \drop
&\approx& {\nainf\over \lambda} t^{1/z-1} \ln {t \over \th} \comma \gap
(z=d+1) \period
                                                                \end{eqarray}
%-----------------------------------------------------------------------
where $F$ here is the hypergeometric function.

Substituting eqs. \eqref{i2-compact}, \eqref{i2-non-compact} and eq.
\eqref{i2-marginal} in eq. \eqref{monday}, and keeping only the
dominant terms, we obtain 
%_______________________________________________________________________
                                                \begin{eq}{chocolate}
\rhoaas(t) 
        \approx 
          \casesshortiii{ \nainf t^{(1/z) -1} / \lambda}  
                        {z <d+1}
                        { \nainf t^{-d/z}}
                        {z > d+1}
                        { \nainf t^{-d/z} \ln (t/\th)}
                        {z =d + 1} 
                                        \comma  (t \rightarrow \infty) \period
                                                                \end{eq}
%-----------------------------------------------------------------------
}% end ignore *************************************************************************************

%*************************************************************************************
%*************************************************************************************
%******************************* END APPENDIX ****************************************
%*************************************************************************************

}

\pagebreak

%****************************** BIBLIOGRAPHY *****************************************

%**********NOTE ! NOTE ! *************************************************************
%***** IT LOOKS FOR A fund_temp.bib FILE !!!!!!!**************************************

%\bibliography{polreaction,allreaction,polymerization,polgeneral,ben,rgcritical,radical,polinterface,general,fund_note}

%*************************************************************************************
%*************************************************************************************
%*************************** BEGIN FIGURE CAPTIONS ***********************************
%*************************************************************************************

                     \begin{thefigures}{99}

\figitem{iface}

Two bulk phases A and B, separated by a thin interface of width $h$,
contain diffusing reactants A and B with densities $\nainf, \nbinf$.
Reactions between A and B molecules, generating inert products, may
occur in the interfacial region only.  The typical distance between
reactants on the B side is $l = a(\nbinf a^d)^{-1/d}$.

During short time 2nd order diffusion-controlled kinetics regimes,
the reaction rate is determined by the small fraction of A-B pairs
which were initially close enough to have diffused and met within time
$t$.  That is, reactions are confined to those pairs whose
exploration volumes (indicated by dashed lines) overlap at time
$t$.  Note that such pairs must be within $\xt$ of the interface.

\figitem{tstar_many}

Schematic of the trajectory of an A particle after time $t$, given
this particle was initially within diffusive range of the interface.
Since the number of encounters with the interface is an increasing
function of time, even for relatively weakly reactive species the A
particle is certain to have reacted at sufficiently long times.  The
timescale is either $\tstarmany$ or $\tl$ (see main text).

\figitem{profile}

A reactant density depletion hole of size $\xt \twid t^{1/z}$ grows
at the interface for long times.  (a) The symmetric case,
$\nainf=\nbinf$.  The reactant density at the interface, $\nas$,
tends asymptotically to zero.  (b) In the asymmetric case,
$\nbinf>\nainf$, the interfacial density $\nas$ on the dilute side
tends to zero, while $\nbs$ asymptotes $\nbinf-\nainf$.

\figitem{sink_terms}

The depletion in the number density of reactive A-B pairs at the
origin from the value it would have in the absence of reactions
originates from three terms, eq. \eqref{exact-rhoabs}.  The two-body
term counts those A-B pairs which would have been at the origin at
time $t$, but failed to arrive because {\em both} members reacted at
an earlier time $t'$ at point $\ratprime$.  The first many-body term
($\Ima$, described in the figure) counts A-B pairs which would have
been at the origin had there been no reactions, but failed to arrive
because {\em one} member of the pair, the A member, reacted at an
earlier time.  The second many body term, $\Imb$, is identical except
the roles of A and B are interchanged.

\figitem{world_lines}

Trajectories of a typical A-B pair which at time $t$ is at the
origin, and whose A member reacted at the interface at time $t'$ with
another B-type particle ($\bar{B}$).  Trajectories are shown projected
onto the $x-\tau$ plane where $x$ is distance from the interface and
$\tau$ is time.  The properties of trajectories of this type
determine the value of the integration which determines the many body
term $\Ima$ (see eq. \eqref{broom} and following discussion in main
text).  At $t'$, the B particle of the pair has $x$-coordinate $\xb'$
which can be (a) in region I, if $\xb'>x_{\tau}$ (as shown in the
figure), or (b) in region II, if $\xb'<x_{\tau}$.

\figitem{fund_phase}

Reaction rate per unit area as a function of time in the $Q$-$\nbinf$
plane ($\nbinf\ge \nainf$).  Units are chosen such that $a =
\ta =1$.  (a) Compact case ($d+1<z$).  (b) Marginal case ($d+1 = z$).
Reaction kinetics in the noncompact case are the same as in the ``weak''
regions of (a) and (b).

\figitem{segregation}

Schematic representation of asymptotic segregation of reactants into
A-rich and B-rich domains of size $\xt$ near to the interface.  This
segregation occurs only for sufficiently low dimensions, $d+1<z$.

                    \end{thefigures}

%*************************************************************************************
%*************************************************************************************
%***************************** END FIGURE CAPTIONS ***********************************
%*************************************************************************************

\pagebreak

%
%
%********************************************************************************
%********************************************************************************
%
%                              FIGURES
%
%********************************************************************************
%********************************************************************************
% NOTE FIGURES ARE EXPECTED IN PORTRAIT MODE IN THE FOLLOWING
%
%\pagestyle{empty}

\begin{figure}[h]

\epsfxsize=\textwidth \epsffile{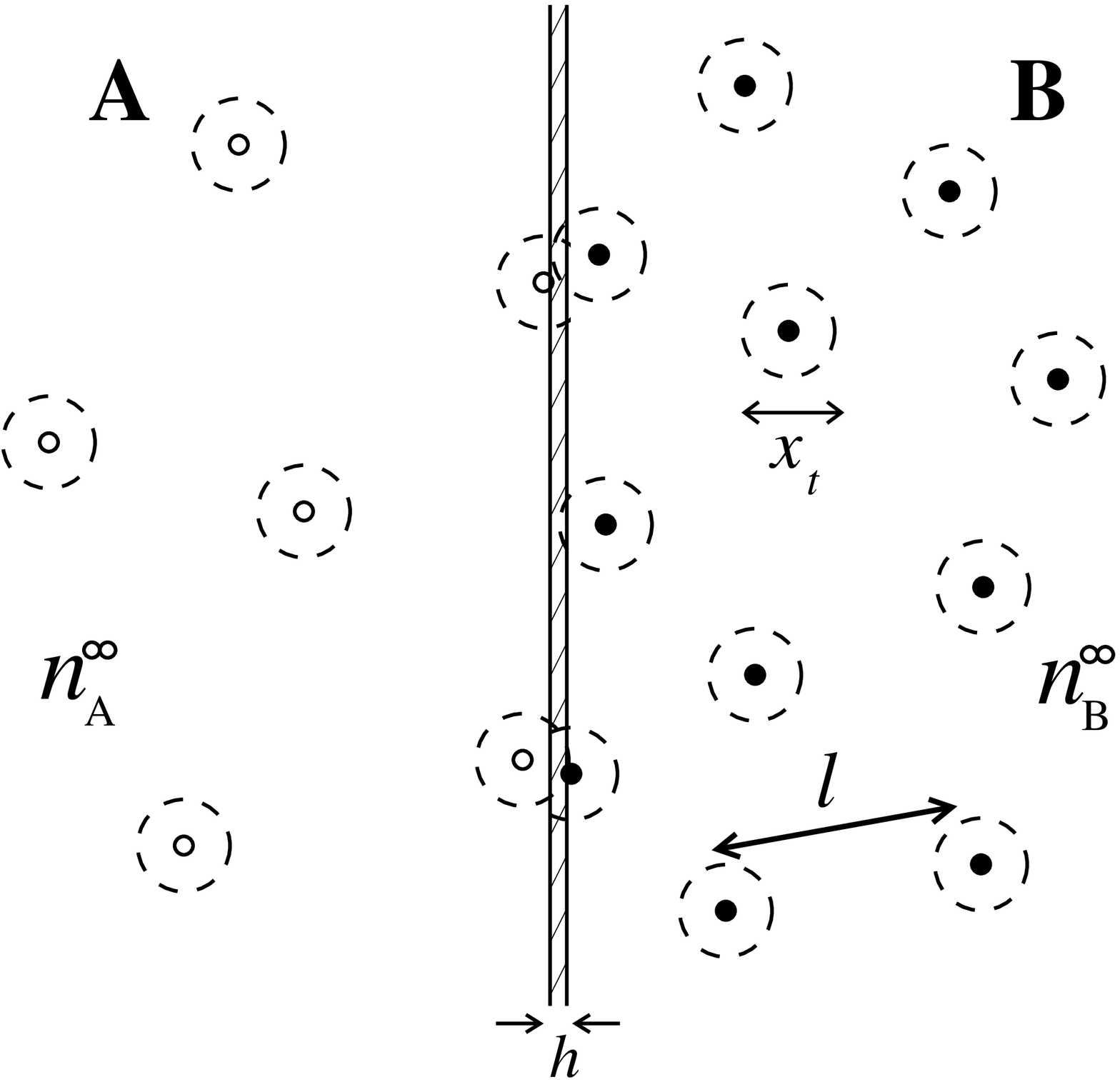}

\end{figure}

\mbox{\ }

\vfill

\addtocounter{fignumber}{1}
\mbox{\ } \hfill {\huge Fig.\@ \thefignumber} 

\pagebreak
%*******************************************************************************
\begin{figure}[h]

\epsfxsize=\textwidth \epsffile{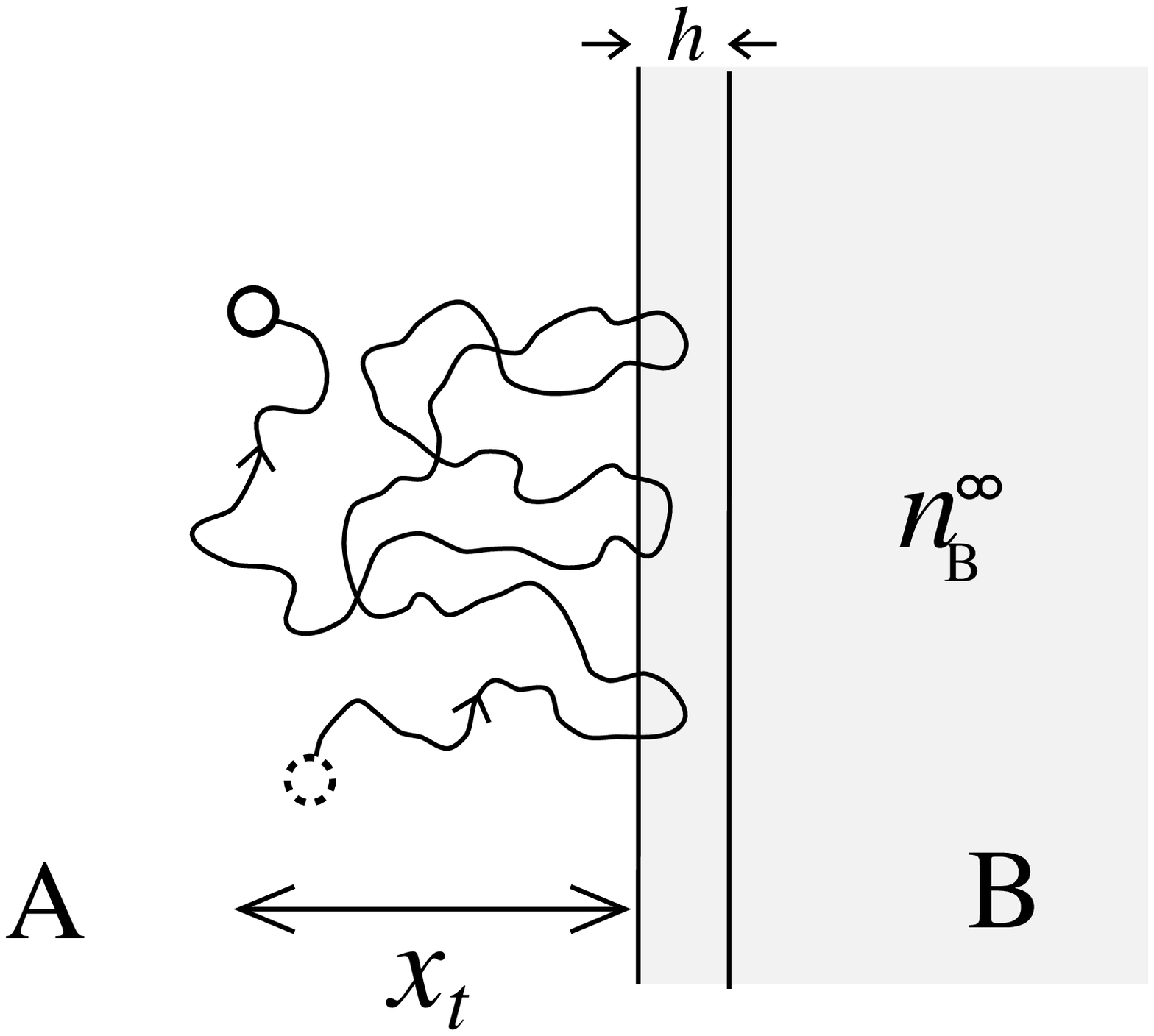}

\end{figure}

\mbox{\ }

\vfill

\addtocounter{fignumber}{1}
\mbox{\ } \hfill {\huge Fig.\@ \thefignumber} 

\pagebreak
%*******************************************************************************
\begin{figure}[h]

\epsfysize=6in \epsffile{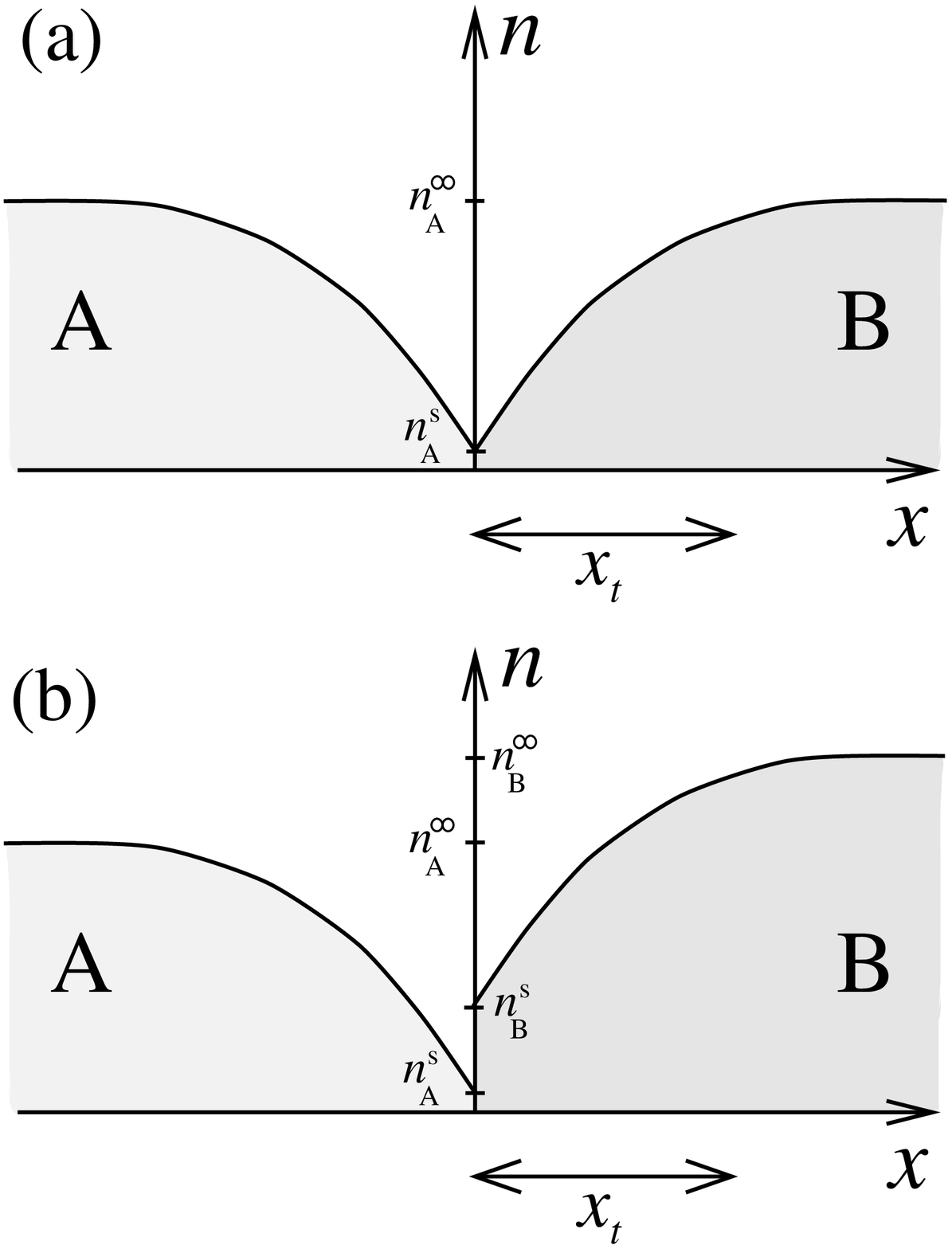}

\end{figure}

\mbox{\ }

\vfill

\addtocounter{fignumber}{1}
\mbox{\ } \hfill {\huge Fig.\@ \thefignumber} 

\pagebreak
%*******************************************************************************

\begin{figure}[h]

\epsfxsize=\textwidth \epsffile{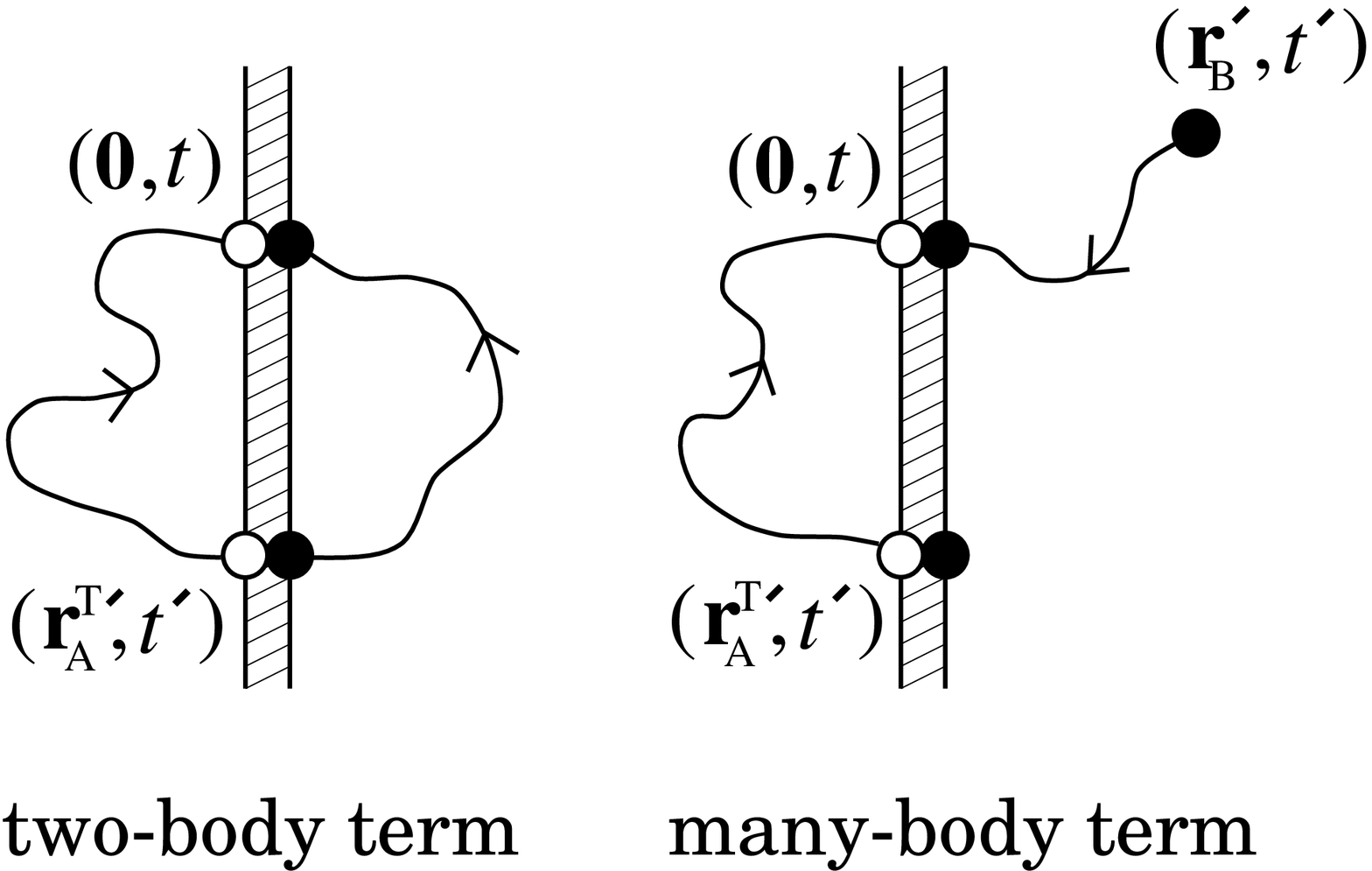}

\end{figure}

\mbox{\ }

\vfill

\addtocounter{fignumber}{1}
\mbox{\ }  \hfill {\huge Fig.\@ \thefignumber} 

\pagebreak

%*************************************************************************************

\begin{figure}[h]

\epsfxsize=\textwidth \epsffile{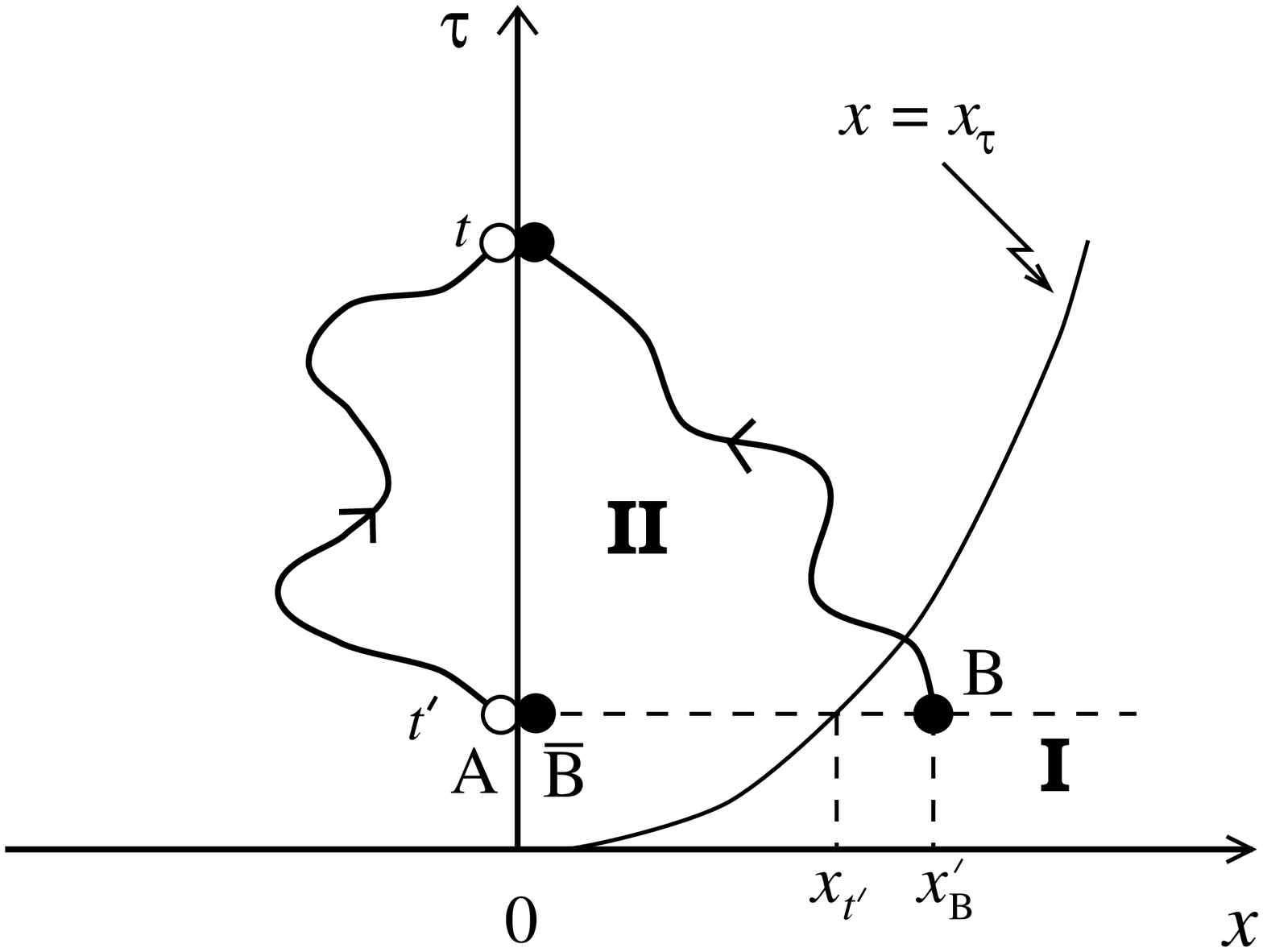}

\end{figure}

\mbox{\ }

\vfill

\addtocounter{fignumber}{1}
\mbox{\ }  \hfill {\huge Fig.\@ \thefignumber} 

\pagebreak

%*************************************************************************************

\begin{figure}[h]

\epsfxsize=\textwidth \epsffile{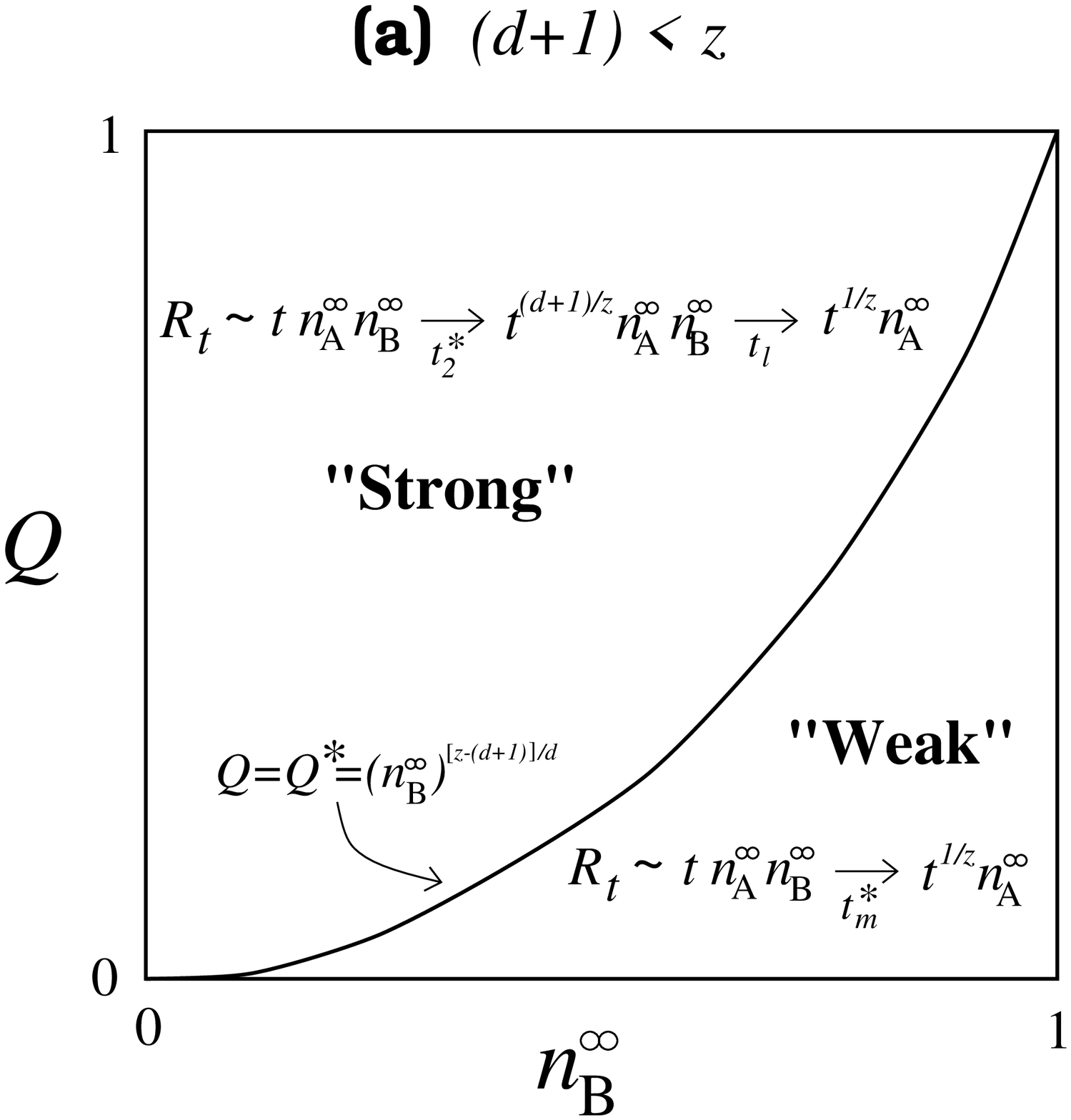}

\end{figure}

\mbox{\ }

\vfill

\addtocounter{fignumber}{1}
\mbox{\ }  \hfill {\huge Fig.\@ \thefignumber (a)} 

\pagebreak

%*************************************************************************************

\begin{figure}[h]

\epsfxsize=\textwidth \epsffile{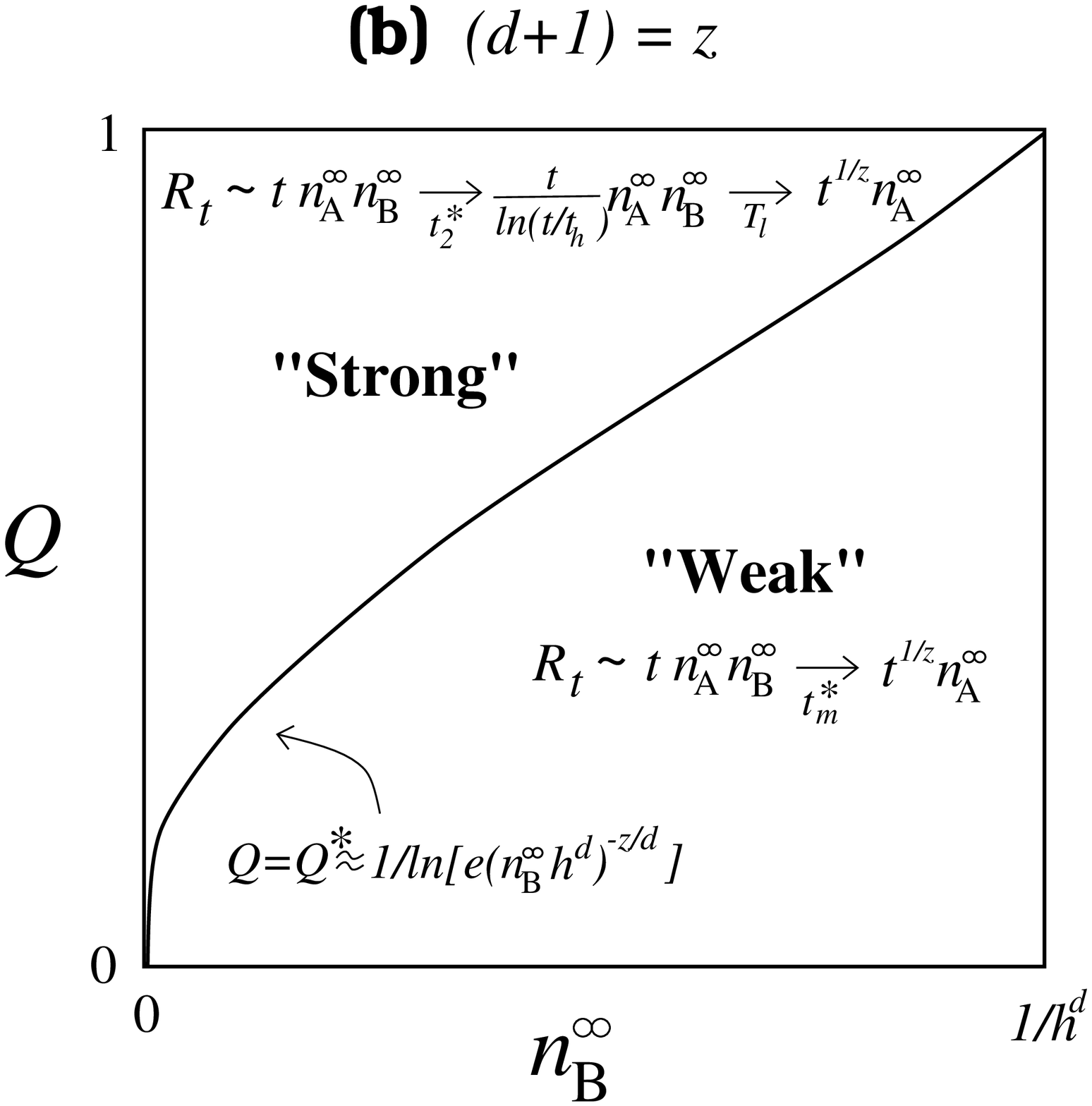}

\end{figure}

\mbox{\ }

\vfill

\mbox{\ }  \hfill {\huge Fig.\@ \thefignumber (b)} 

\pagebreak

%*************************************************************************************

\begin{figure}[h]

\epsfxsize=\textwidth \epsffile{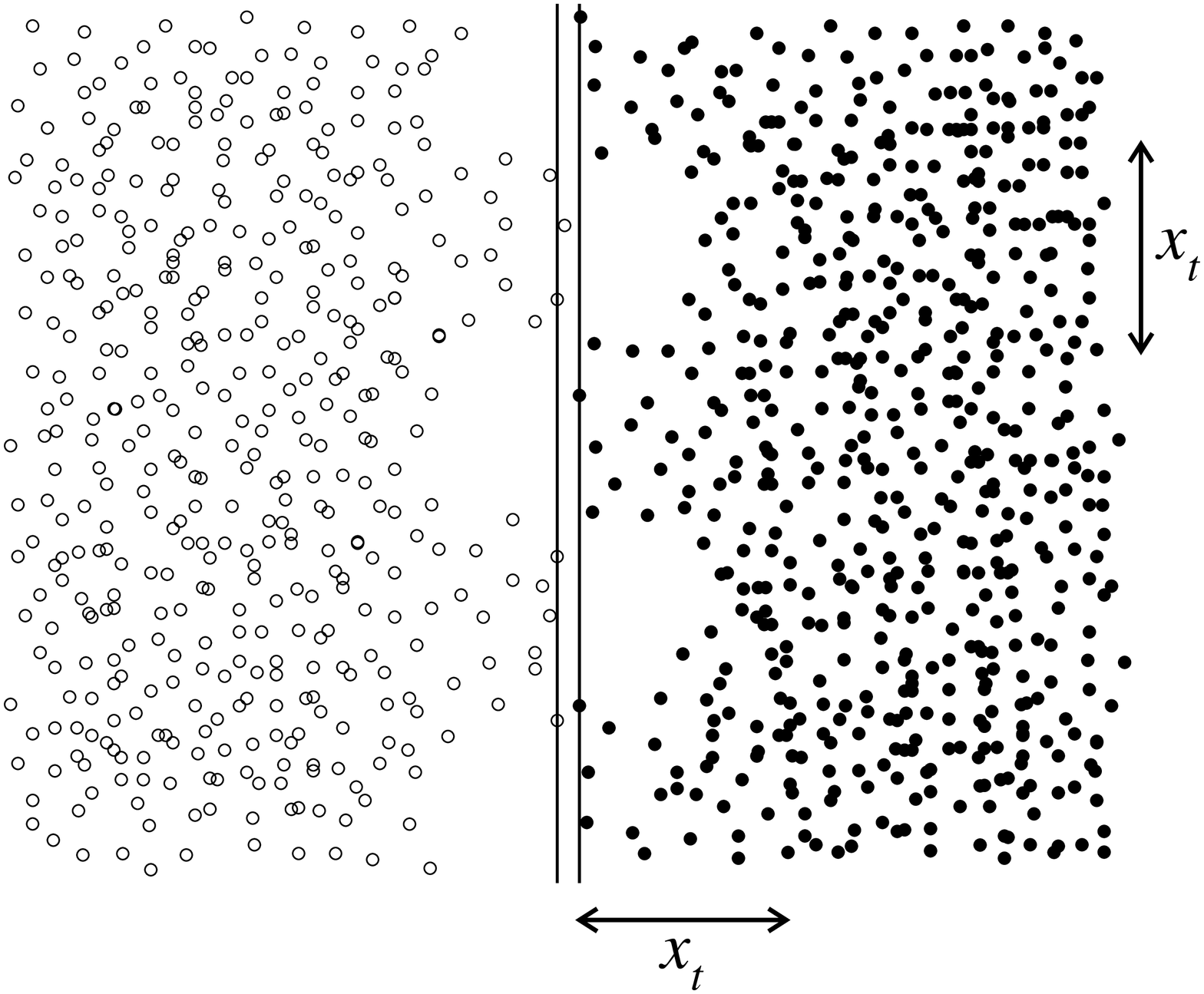}

\end{figure}

\mbox{\ }

\vfill

\addtocounter{fignumber}{1}
\mbox{\ }  \hfill {\huge Fig.\@ \thefignumber} 

\pagebreak

%*************************************************************************************

%*************************************************************************************
%*************************************************************************************
%***************************** END FIGURES *******************************************
%*************************************************************************************
%*************************************************************************************
%*************************************************************************************

\end{document}